\pdfoutput=1
\documentclass[12pt, oneside, a4paper]{article}
\usepackage{enumitem}
\usepackage{pgfplots}
\pgfplotsset{compat=1.18}
\usepackage{amsmath}
\usepackage{amssymb}
\usepackage{amsthm}
\usepackage{bbm}
\usepackage{amsfonts}
\usepackage{mathtools}
\usepackage{tikz-cd}
\usepackage{booktabs}
\usepackage{multirow}
\usepackage{algorithm}
\usepackage{algpseudocode}
\usepackage{xcolor}
\usetikzlibrary{shapes,arrows}
\mathtoolsset{showonlyrefs}

\usepackage{graphicx}
\usetikzlibrary{spy, calc}

\DeclareMathOperator*{\argmin}{arg\,min}

\usepackage{etoolbox}

\usepackage[nottoc,notlot,notlof]{tocbibind}

\usepackage{listings}
\usepackage{color}
\definecolor{mygreen}{RGB}{28,172,0}
\definecolor{mylilas}{RGB}{170,55,241}

\newtheorem*{theorem*}{Theorem}
\newtheorem*{remark*}{Remark}

\usepackage{amsthm}
\theoremstyle{remark}

\usepackage{xpatch}

\makeatletter
\xpatchcmd{\@thm}{\thm@headpunct{.}}{\thm@headpunct{}}{}{}
\makeatother

\usepackage{hyperref}
\usepackage[british]{babel}
\usepackage{caption}
\usepackage{subcaption}
\usepackage{graphicx}
\usepackage[justification=centering]{caption}
\usepackage{stackengine}
\usepackage{float}
\usepackage{longtable}
\usepackage{tabularx}
\newcolumntype{Y}{>{\centering\arraybackslash}X}
\usepackage{mathtools}
\usepackage{enumitem}
\usepackage{titlesec}
\titleformat{\chapter}[display]
{\normalfont\huge\bfseries}{}{0pt}{\Huge}
\titlespacing*{\chapter} {0pt}{20pt}{40pt}
\usepackage[thinlines]{easytable}
\usepackage{array}

\usepackage[toc]{appendix}

\usepackage{authblk}
\usepackage{graphicx}

\usepackage[nocompress]{cite}

\usepackage[left=3cm, right=3cm, top=3.5cm, bottom=3.5cm]{geometry}

\title{Diffusion Priors for Plasma Tomography: Bayesian Data-Driven Methods Applied to Soft X-Ray Reconstruction}
\author[1]{D. Hamm\thanks{These authors contributed equally to this work. Their names are sorted alphabetically.\\$\phantom{1111}$Corresponding authors: daniele.hamm@epfl.ch, youssef.haouchat@epfl.ch}${^{,}}$}
\author[2]{Y. Haouchat${}^{\,*,}$}
\author[1]{C. Theiler}
\affil[1]{EPFL - Swiss Plasma Center, CH-1015 Lausanne, Switzerland}
\affil[2]{EPFL - Biomedical Imaging Group, CH-1015 Lausanne, Switzerland}
\date{}
\date{}

\begin{document}
\maketitle

\newcommand{\vx}{\mathbf{x}}
\newcommand{\vmu}{\boldsymbol{\mu}}
\newcommand{\vsigma}{\boldsymbol{\sigma}}

\begin{abstract}
Plasma emissivity reconstruction from tomographic data is typically a severely ill-posed problem.
Established approaches rely on penalized likelihood formulations (including Bayesian ones) or, more recently, on deep learning.
The former are robust, but their performance is limited by the expressiveness of the employed regularization/prior.
The latter can reduce computation time, but are typically tied to the diagnostic configuration used for training, and uncertainty quantification is usually not available.
We propose using diffusion models to enhance the expressiveness of the prior adopted for plasma tomographic reconstruction.
We learn the prior by training a neural network on realistic synthetic data, then use the DiffPIR algorithm to perform
posterior sampling, combining the prior with the likelihood corresponding to a given reconstruction problem.
As learning and reconstruction are decoupled, the learned prior is agnostic to the diagnostic configuration.
We conduct a phantom-based study, applying DiffPIR to three soft X-ray diagnostics of the TCV tokamak: RADCAM's SXR, the planned DMPX upgrade, and Pilatus.
For SXR and DMPX, we show that DiffPIR, despite not relying on the magnetic equilibrium, can outperform an equilibrium-informed Bayesian approach that anisotropically promotes smoothness along flux surfaces.
For the single-camera Pilatus, we propose DiffRD, an equilibrium-informed modification of DiffPIR that outperforms the other considered methods.
We show that uncertainty quantification based on the variability of the posterior samples is more informative about the localization of reconstruction errors than that of the anisotropic Bayesian baseline.
We also discuss the sensitivity of the learned prior to out-of-distribution data.
Finally, we apply DiffPIR to experimental SXR data; the obtained reconstructions are found to be consistent with the magnetic equilibrium, even though the equilibrium does not enter the reconstruction.

\end{abstract}

\newpage

\section{Introduction}
\label{introduction}
Fusion devices often feature diagnostics that measure the power radiated by the plasma over different energy ranges (e.g., soft and hard x-ray systems, visible light cameras, broadband bolometer detectors).
Tomography can then be used to reconstruct plasma emissivity profiles.
Rather than attempting to reconstruct 3D radiation patterns, plasma imaging techniques typically reconstruct 2D images corresponding to the radiation profiles on poloidal cross-sections of the devices, relying on the assumption of toroidally symmetric emissivity \cite{ingesson_chapter_2008}.
Despite this simplifying assumption, plasma tomographic reconstruction remains a challenging problem due to the extremely sparse-view and limited-angle nature of the available measurements.
Solving the resulting underdetermined and ill-conditioned tomographic inverse problem requires incorporating some form of prior knowledge about the emissivity field.
Often, knowledge of the magnetic equilibrium is exploited to promote smoothness along rather than across flux surfaces, justified by the strong anisotropy of transport in magnetized plasmas.
\\
The most widely employed techniques rely on penalized likelihood formulations that can be described within a unifying Bayesian framework \cite{hamm2025}.
Most of these approaches use Gaussian noise models that yield Gaussian likelihoods \cite{hamm2025}.
Prior knowledge is then modeled with generalized Tikhonov approaches based on gradient or Laplacian smoothness terms \cite{anton_x-ray_1996, ingesson_chapter_2008, hamm2025}, sometimes inspired by Minimum Fisher Information principles \cite{anton_x-ray_1996, odstrcil_modern_2012}.
Alternatively, kernels that model the correlation between neighboring pixels can be used to define Gaussian priors, leading to the popular Gaussian Process Tomography framework \cite{li_bayesian_2013, moser_gaussian_2022}.
All these approaches allow \cite{hamm2025} Bayesian uncertainty quantification based on the posterior standard deviation, estimated either by posterior sampling or, in the case of Gaussian posteriors, with closed-form expressions.
The Maximum Likelihood method \cite{craciunescu_maximum_2008, craciunescu_maximum_2018, wyss2026} follows a different approach: it adopts a Poisson likelihood model and introduces prior knowledge implicitly, by interleaving smoothing operations between successive likelihood maximization steps.
The method implements \cite{craciunescu_maximum_2018, wyss2026} a form of frequentist uncertainty quantification that estimates the sensitivity of the reconstruction to the measurement noise.
All the methods discussed in this paragraph allow the optional implementation of priors promoting anisotropic smoothness.\\
These methods have been successfully used for many years for plasma tomographic reconstruction, due to their robustness and interpretability.
However, when the posterior is Gaussian, the quantified uncertainty is predominantly determined by the interplay between the prior and likelihood (with only indirect dependence on the observed measurements) \cite{hamm2025}, while in the Maximum Likelihood case it is highly correlated \cite{craciunescu_maximum_2018, wyss2026} to the reconstruction itself.
In both cases, although potentially useful to define operative credible/confidence intervals, the estimated uncertainty is not significantly informative about the localization of the actual reconstruction error.
Moreover, anisotropic approaches have the downsides of attenuating emissivity features that are not flux-aligned, and of propagating into their estimates any errors in the magnetic reconstruction.
This could limit their applicability in future DEMO reactors, as in-vessel magnetic probes are expected to degrade under reactor-relevant conditions, while ex-vessel backups are less accurate \cite{biel2019}.
Finally and crucially, the expressiveness of these priors is limited by their simple functional form, which cannot accurately model the complex processes determining the plasma emissivity, even in the most sophisticated formulations accounting for region-dependent anisotropic smoothness.
For all these reasons, it is important to explore alternative approaches.\\
Deep learning approaches to plasma tomographic reconstruction have also been proposed.
The first attempts were based on end-to-end convolutional neural networks (CNNs) that learn a direct mapping from the tomographic measurements to the 2D emissivity profile \cite{matos_2017, ferreira_deep_2019}.
Training relies on a set of reconstructions obtained from traditional penalized likelihood methods, of which the learned model is thus a surrogate.
Although fast, these approaches provide no uncertainty quantification and inherit the regularization of the reconstructions they were trained to reproduce, without increasing expressiveness.
Moreover, they are trained for a specific diagnostic layout: retraining is needed if the diagnostic configuration changes, and robustness to detector failures is a concern.
Other approaches are instead trained on synthetic emissivity profiles with known ground truth \cite{Jardin_2024}, avoiding inherited regularization but still providing no uncertainty quantification and remaining tied to a specific diagnostic layout.\\
Another line of deep learning work makes the physics explicit rather than implicit, either by training a CNN to select the most appropriate model among a list of predefined Gaussian priors \cite{matos_deep_2020}, or by embedding the forward model and suitably defined PDE residuals into a physics-informed loss used to train a fully-connected neural network \cite{Rossi_2023, Rossi_2025}.
This second approach, based on physics-informed neural networks (PINNs) \cite{raissi_2019}, can also be used \cite{Rossi_2025} to promote emissivity profiles that vary smoothly also in time rather than only in space: a promising strategy that remains little explored in plasma tomography and that connects to the literature on penalized likelihood approaches combining space and time regularization \cite{Schmitt_2002}.
PINN-based approaches require a separate training for each analyzed time slice or time window, which considerably increases the computational cost of inference; accordingly, they are proposed in \cite{Rossi_2025} as offline tools for high-accuracy, super-resolved analysis of selected time windows.
Regarding uncertainty quantification, \cite{matos_deep_2020} inherits the limitations of Gaussian posteriors discussed above;
for PINNs, it is achievable through Bayesian formulations \cite{yang_bpinn_2021}, but is yet to be explored for plasma tomography \cite{Rossi_2023, Rossi_2025} and would further increase the computational cost.
Finally, in both \cite{matos_deep_2020} and \cite{Rossi_2023, Rossi_2025}, the prior
is hand-designed: it is either Gaussian \cite{matos_deep_2020} or based on anisotropic diffusion equations similar to those used in \cite{hamm2025, wyss2026}, and only its hyperparameters are learned.
Once again, the expressiveness of the prior is not significantly increased.\\
In this paper, we propose to use \emph{diffusion models} for plasma tomographic reconstruction, to address some of the weaknesses of the methods discussed in the previous paragraphs.
In simple terms, diffusion models work as follows.
If we start from a clean image (it might be a photograph, a painting, or a plasma emissivity profile) and iteratively add a small amount of Gaussian noise to it, we eventually converge to an image which is pure noise and that does not retain any of the features of the original image.
By learning to reverse all the noise addition steps in this dynamic, we obtain a \emph{generative denoiser} able to generate clean images from pure noise realizations.
Diffusion models achieve this by training on a set of clean images a neural network that effectively learns the gradient of noise-perturbed versions of the log-prior, across a range of  noise levels.
The learned gradients, starting from a random initial condition, are used to guide the sampling process towards regions of high probability density.
If properly trained, the model generates clean images that are novel rather than memorized copies of the training samples.
\\
Diffusion models (see \cite{daras_2024} for a review) have become extremely popular in recent years.
The possibility of using them to generate high quality samples was demonstrated by the Denoising Diffusion Probabilistic Models (DDPM) framework \cite{ho_2020}.
These models can be viewed as discretizations of the underlying time-continuous reverse denoising dynamics \cite{song_2021}.
Since then, they have been shown \cite{dhariwal_2021} to produce higher quality samples than Generative Adversarial Networks (GANs), which were the previous state of the art in generative modeling.
They have become known to the general public through the popular Stable Diffusion \cite{rombach_2022} and OpenAI's DALL$\cdot$E 2 \cite{ramesh_2022} models.
Diffusion models have been shown to achieve excellent performance also in scientific imaging applications \cite{song_2022}.
Several approaches have been proposed for approximate sampling from the posterior distributions arising in general noisy inverse problems \cite{zhu_2023, chung_2024}.
In this work, we rely on the Denoising Diffusion Models for Plug-and-Play Image Restoration (DiffPIR) approach \cite{zhu_2023}.
DiffPIR is inspired by the Plug-and-Play framework, which first showed that denoisers could be combined with physics-based likelihood models to solve general inverse problems \cite{venka2013}; we refer the interested reader to \cite{kamilov2023} for a review.
\\
We show that diffusion models lead to more expressive priors and better uncertainty quantification than the plasma tomography methods discussed earlier in the section.
In terms of inference time, they are slower than end-to-end pre-trained neural network approaches (which, however, are tied to a specific diagnostic), and comparable to traditional iterative techniques: for a given set of noisy measurements, $100$ posterior samples can be drawn in a few seconds.
They are more time-efficient than PINN approaches, as they decouple the learning from the reconstruction phase: the model can be readily plugged into an arbitrary reconstruction problem.
We find that the trained model, after setting its hyperparameters once for a given diagnostic,
achieves near-optimal reconstruction performance without further sample-specific hyperparameter tuning.
This removes a cost that, in practice, slows down penalized likelihood approaches, in particular the otherwise efficient Gaussian Process Tomography.
We opt to train the model on a set of physically realistic emissivity phantoms, so that regularization is not inherited from the reconstructions computed with other methods.
This ensures that the expressiveness of the learned prior is determined only by the quality and variability of the training samples, avoiding learning artifacts induced by other reconstruction codes.
We will also discuss the inevitable limitation that diffusion models share with all data-driven approaches, i.e., their performance decrease when confronted with severely out-of-distribution data that are considerably different from the training samples: comparing the computed reconstructions to simpler benchmarks allows identifying these events.
To the best of our knowledge, diffusion models have never been applied to plasma tomography, and little literature exists on their application to problems as severely underdetermined as the ones here investigated, where a few hundred measurements at most constrain several thousand pixels.\\
We apply diffusion models to soft X-ray tomographic reconstruction.
We consider three soft X-ray diagnostics from the TCV tokamak \cite{Theiler_2026}: RADCAM's SXR system \cite{sheikh_radcam-radiation_2022}, the imminent upgrade \cite{chellai_2026} of the DMPX multi-wire system \cite{dreval_2023}, and the recently installed \cite{chellai_2025} Pilatus multi-energy diagnostic \cite{chellai_2024}.
After a single training run of a few hours, the learned diffusion prior is ready to be deployed.
As the learned prior is fully agnostic to the target soft X-ray reconstruction problem on TCV, it can be directly used for tomographic reconstruction from all three diagnostics, without any retraining needed. Transferring the model to another device, instead, would likely require retraining.
\\
    Our main contributions in this paper are as follows.
    \begin{itemize}
    \item We propose a Bayesian data-driven framework, based on diffusion models, for tomographic reconstruction of plasma emissivity profiles.
    We describe the algorithms that can be used, within this framework, for prior and posterior sampling.
    \item We show that, for diagnostics featuring multiple cameras such as SXR and DMPX, the equilibrium-free DiffPIR can lead to better reconstruction and more meaningful uncertainty quantification than equilibrium-informed penalized likelihood approaches such as the reaction-diffusion (RD) one from \cite{hamm2025}.
    \item We propose DiffRD, an equilibrium-informed DiffPIR variant that integrates the RD prior into the sampling procedure. We demonstrate, for the single-camera Pilatus diagnostic, that DiffRD can outperform both RD and DiffPIR.
    \item We provide open access to the implementation of all the algorithms discussed in this work. The presented results can be reproduced by following the instructions reported at the paper repository \url{https://github.com/HaouchatY/plasma_diffusion.git}.
    \end{itemize}
The rest of this article is structured as follows.
In Section \ref{inverse_problems_tomography}, we introduce the three TCV soft X-ray diagnostics considered in this work (SXR, DMPX, Pilatus).
We then recall the RD Bayesian framework used as baseline throughout the paper, describing its (stochastic) gradient flow approach for MAP estimation and posterior sampling.
In Section \ref{diffusion_models}, we provide an introduction to diffusion models, describing the DDPM and DiffPIR algorithms for prior and posterior sampling, respectively.
Moreover, we propose an equilibrium-informed version of the DiffPIR algorithm.
We then consider the application of these algorithms to soft X-ray imaging.
We show how to learn expressive diffusion priors from realistic synthetic data, and how to use these priors for tomographic reconstruction from arbitrary soft X-ray diagnostics.
In Section \ref{results}, we compare the diffusion and RD frameworks.
First, we report the results of phantom-based studies on the three diagnostics, discussing the improvements enabled by the diffusion framework in terms of reconstruction quality and of ability to provide meaningful uncertainty quantification.
Then, we use the DiffPIR algorithm for reconstruction from experimental data from the SXR diagnostic, showing that the proposed methods enable experimentally relevant inference.
We conclude the section by discussing the out-of-distribution limitations of the diffusion framework.
Finally, in Section \ref{summary_and_discussion}, we summarize the obtained results and discuss possible improvements and extensions of our work.

\section{Soft X-Ray Diagnostics and Baseline Model}
\label{inverse_problems_tomography}
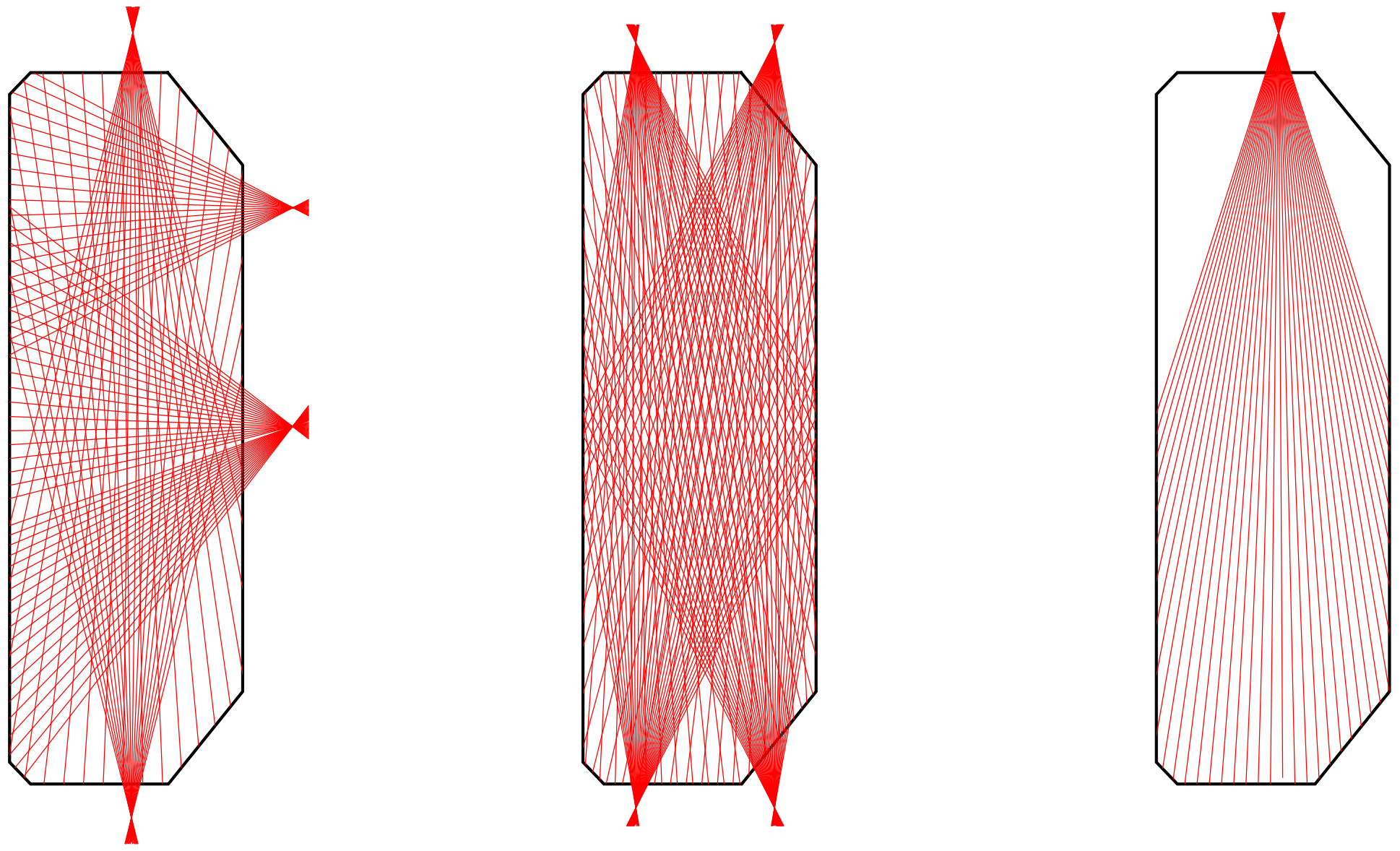
\begin{figure}[t!]
          \centering
          \begin{tikzpicture}
            \node[inner sep=0pt] (sxr) at (0,0)
              {\includegraphics[width=0.22\textwidth]{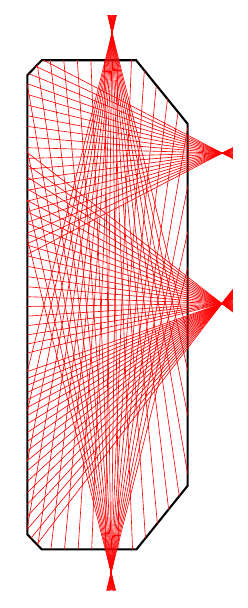}};
            \node[inner sep=0pt] (dmpx) at (0.35\textwidth,0)
              {\includegraphics[width=0.22\textwidth]{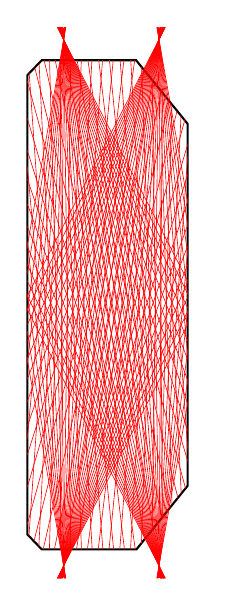}};
            \node[inner sep=0pt] (pilatus) at (0.7\textwidth,0)
              {\includegraphics[width=0.22\textwidth]{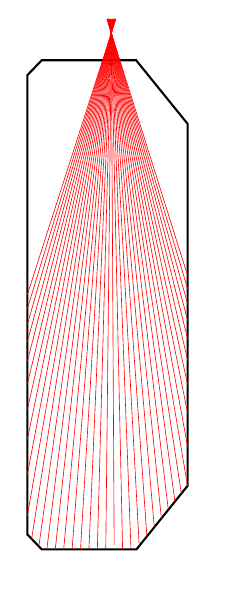}};

            \node[below=2mm] at (sxr.south)  {\small (a) SXR};
            \node[below=2mm] at (dmpx.south) {\small (b) DMPX};
            \node[below=2mm] at (pilatus.south)   {\small (c) Pilatus};
          \end{tikzpicture}
          \captionsetup{justification=justified,singlelinecheck=false}
          \caption{Poloidal cross-sections of TCV with line-of-sight configurations of the three considered soft X-ray diagnostics. In Fig.\ref{fig:los_configurations}a-\ref{fig:los_configurations}c, the RADCAM's SXR system, the planned DMPX upgrade, and the Pilatus diagnostic, respectively. While all the lines-of-sight of the SXR system are shown, only every other line-of-sight is shown  for DMPX and Pilatus, for visualization clarity.}
          \label{fig:los_configurations}
\end{figure}

In this work, we consider tomographic reconstruction from RADCAM's SXR diodes, the planned DMPX upgrade, and the Pilatus detector. These three diagnostics, which are all pinhole or slit cameras that measure soft X-ray radiation from the plasma, are shown in Fig.\ref{fig:los_configurations}.
The RADCAM's SXR system \cite{sheikh_radcam-radiation_2022}, that we will simply refer to as SXR from here on, is composed of $100$ photodiodes divided in five cameras of $20$ detectors.
It was designed to optimize the coverage in the poloidal plane of the TCV tokamak, with rather large beam widths but minimal overlap between adjacent detectors.
It allows tomographic reconstruction of the soft X-ray emissivity at up to $100\,\mathrm{kHz}$, ideal for the study of MHD activity.
The new
DMPX \cite{chellai_2026}
is composed of four cameras,
each composed of two superimposed chambers to allow simultaneous acquisition over two energy ranges.
As each chamber is made of $64$ wires, there are $256$ total detectors for each energy range.
The diagnostic was designed to ensure a $100\,\mathrm{kHz}$ acquisition frequency and fine spatial resolution.
While the measurements of each chamber allow 2D emissivity reconstruction in the respective energy ranges, the ratio of the two can be used to reconstruct the electron temperature $T_e$.
The single-camera Pilatus \cite{chellai_2025} provides energy-resolved soft X-ray measurements at $500\,\mathrm{Hz}$, with high spatial resolution.
Different energy thresholds can be prescribed for each of the $\approx 10^5$ pixels composing the detector, allowing great flexibility in terms of spatial and energy resolution.
In its current configuration, the individual pixels are grouped to form $6$ groups of $80$ larger rectangular pixels, with each group set to a different threshold.
As for DMPX, both soft X-ray and $T_e$ profiles can be inferred.
Due to the single-camera layout, only 1D reconstruction of the radial profiles is typically attempted, constraining the emissivity to be constant on flux surfaces.\\
In the following, for simplicity, we will only refer to soft X-ray emissivity reconstruction; the discussion directly applies to $T_e$ reconstruction from soft X-ray brightness ratios.
Moreover, we study only one of the two DMPX chambers, and a single energy level for Pilatus.
We thus effectively consider SXR, DMPX and Pilatus diagnostics made of $M=100,256,80$ detectors, respectively.
Let us consider 2D reconstruction on a discrete grid composed of $N=N_z\times N_r$ rectangular pixels, where $N_z=120$ and $N_r=40$ are the number of pixels  used to discretize the poloidal plane in the vertical and radial directions.
The problem of reconstructing a 2D emissivity profile $\mathbf{x}\in\mathbb{R}^N$ from the tomographic data $\mathbf{y}\in\mathbb{R}^M$ can be formalized as the linear inverse problem
\begin{equation}\label{ip_data}
    \mathbf{y} = \mathbf{T}\mathbf{x} + \boldsymbol{\varepsilon},
\end{equation}
where $\boldsymbol{\varepsilon}\in\mathbb{R}^M$ is the noise affecting the measurements and $\mathbf{T}\in\mathbb{R}^{M\times N}$ is the forward model describing the diagnostic, referred to as a geometry matrix in the plasma imaging community.
For all three diagnostics, we compute physically accurate geometry matrices using the voxel-to-detector approach described in \cite{hamm2026}, taking into account the full 3D geometry of the imaging system.\\
In a Bayesian framework,
inference is based on the posterior distribution $p(\mathbf{x}\vert \mathbf{y})$, which describes our knowledge about $\mathbf{x}$ given the observed measurements $\mathbf{y}$.
By Bayes' rule, the posterior satisfies
\begin{equation}\label{bayes_formula}
    p(\mathbf{x}\vert \mathbf{y}) \propto p(\mathbf{y} \vert \mathbf{x}) p(\mathbf{x})\,,
\end{equation}
where $p(\mathbf{y} \vert \mathbf{x})$ is the likelihood of measuring $\mathbf{y}$ for a given $\mathbf{x}$, and $p(\mathbf{x})$ is the prior distribution that models the assumed properties of the emissivity profile.
The most probable solution to the inverse problem, i.e., the Maximum A Posteriori (MAP) estimator $\mathbf{x}_{_{MAP}}$, can be computed by maximizing the posterior in Eq.\eqref{bayes_formula}.
While typically cheap to compute, the MAP only provides a point estimate: richer information, such as uncertainty quantification, can be extracted by sampling from the posterior by means of Markov Chain Monte Carlo (MCMC) methods.
If an explicit expression of the \emph{score} of the prior $\nabla \log p(\mathbf{x})$ is known, gradient-based schemes can be used for inference.
In particular, as discussed in \cite{hamm2025}, the MAP can be computed relying on a gradient ascent algorithm, while posterior sampling can be performed by means of the Unadjusted Langevin Algorithm (ULA) \cite{roberts_exponential_1996, durmus_nonasymptotic_2017, durmus_high-dimensional_2019}.
ULA, which is the Euler–Maruyama discretization of the overdamped Langevin stochastic differential equation \cite{pavliotis_stochastic_2014}, acts as the stochastic counterpart of the gradient flow used for MAP estimation.
The resulting discretized (stochastic) gradient flow algorithm then reads
\begin{equation}\label{GDULA}
   \mathbf{x}_{k+1} = \underbrace{\underbrace{\mathbf{x}_k + \tau \nabla \log p(\mathbf{y}\vert\mathbf{x}_k) + \tau \nabla\log p(\mathbf{x}_k)}_{\text{MAP computation: gradient ascent}} \;\Big(+ \sqrt{2\tau}\,\mathbf{z}_k\Big)}_{\text{MCMC sampling: Unadjusted Langevin Algorithm (ULA)}},\;\; \mathbf{x}_0\!=\!\bar{\mathbf{x}}_0,\;\; k\geq0,
\end{equation}
where $(\mathbf{z}_k)_{k\geq0}\sim\mathcal{N}(\mathbf{0}, \mathbf{I})$ are independent realizations of standard $N$-dimensional Gaussian vectors.
In Eq.\eqref{GDULA}, the randomness introduced by the stochastic term $\sqrt{2\tau}\,\mathbf{z}_k$ allows the exploration of the distribution by the sampler, with the $\sqrt{2\tau}$ scaling ensuring that the chain admits the posterior as its invariant measure in the limit $\tau\to0$. For fixed $\tau>0$ the chain targets the posterior only approximately.
\\
In this work, we use the reaction-diffusion (RD) Bayesian framework of \cite{hamm2025} as our baseline.
In particular, we consider a Gaussian noise model $\varepsilon_m\sim\mathcal{N}(0,\eta^2)$, with constant noise level $\eta$ for all detectors; detector-dependent noise levels could also be assumed.
The chosen noise model yields the Gaussian likelihood
\begin{equation}\label{gaussian_likelihood}
    \mathbf{y}\vert\mathbf{x} \sim \mathcal{N}(\mathbf{T}\mathbf{x}, \eta^2\mathbf{I})\;\;
    \implies \;\; p(\mathbf{y}\,\vert\,\mathbf{x}) \propto \exp\Big(-\frac{(\mathbf{y}-\mathbf{T}\mathbf{x})^T(\mathbf{y}-\mathbf{T}\mathbf{x})}{2\eta^2}\Big).
\end{equation}
Moreover, we consider the equilibrium-informed anisotropic diffusion prior
\begin{equation}\label{eq_diffusion_prior}
p(\mathbf{x})\propto \exp\!\!\,\Big(\!\!-\frac{\lambda}{2} \big\Vert \sqrt{\mathbf{D}(\psi,\alpha)}\boldsymbol{\nabla}\mathbf{x} \big\Vert_2^2 \Big)\;,%
\end{equation}
where $\lambda>0$ is a prior hyperparameter that acts as regularization
parameter, and $\mathbf{D}(\psi,\alpha)$ is a diffusion tensor with anisotropic parameter $\alpha$, promoting smoothness along the isolines of the poloidal flux function $\psi$.
As shown in \cite{hamm2025}, the MAP gradient flow arising from Eqs.\eqref{gaussian_likelihood} and \eqref{eq_diffusion_prior} can be interpreted as a reaction-diffusion differential equation, hence the RD label, which we will use exclusively from here on.
As discussed in \cite{hamm2025}, the resulting negative log-posterior $-\log p(\mathbf{x}\vert\mathbf{y})$ is (strictly) convex with Lipschitz-continuous gradient.
In the baseline reconstruction model, we additionally introduce a positivity constraint, handled via proximal techniques, see \cite{hamm2025}.
Moreover, unless otherwise stated, only the core region is reconstructed, assuming that soft X-ray emission is localized inside it.
\\
The prior score $\nabla \log p(\mathbf{x})$ in Eq.\eqref{GDULA} is analytically known for the hand-designed priors discussed in Section \ref{introduction}.
If representative examples (phantoms) of the emissivity profiles to be reconstructed are known, however, the score can be learned.
This is precisely what diffusion models do, as we describe next.

\newpage

\section{Diffusion Models for Soft X-ray Imaging}
\label{diffusion_models}
The ULA recursion in Eq.\eqref{GDULA} provides samples from a target probability distribution $p$ whenever its score is known.
In high dimensions, ULA is computationally efficient whenever $-\log p$ is convex with Lipschitz-continuous gradient \cite{durmus_efficient_2018, durmus_nonasymptotic_2017}.
This is the case of the RD model recalled in Section \ref{inverse_problems_tomography}; more generally, the condition is often verified by imaging models featuring hand-designed priors \cite{hamm2025}.
For more complex target distributions, however, ULA can converge slowly: when the target concentrates its mass around well-separated modes or a low-dimensional set, the chain needs many iterations to reach and explore the relevant regions \cite{song_2019}.
A remedy is to let the target change along the iterations, implementing a so-called \emph{annealing} strategy \cite{kirkpatrick1983, neal2001ais}: the resulting \emph{time-inhomogeneous} variant of ULA reads
\begin{equation}\label{annealed_ula}
    \mathbf{x}_{k+1} = \mathbf{x}_k + \tau_k \nabla\log p_{t_k}(\mathbf{x}_k) + \sqrt{2\tau_k}\,\mathbf{z}_k\;,\quad k\geq0\;,
\end{equation}
where $\tau_k$ is now a time-dependent step size and $(p_{t_k})_{k\geq0}$ is a prescribed sequence of densities that ends at the target $p$.
The sequence typically starts from a heavily smoothed version of $p$, with each successive density less smoothed than the previous one.
The early iterations thus explore a simple landscape in which large moves are possible, while the late iterations refine the samples on the target itself.
Diffusion models implement precisely this scheme \cite{song_2019}: they adopt a specific smoothing sequence, whose scores are learned from data.\\
In this section, we present diffusion models and show how to use them to learn a prior for soft X-ray imaging (Section \ref{prior_sampling}), then describe the posterior sampling methods that we rely on in this paper for tomographic reconstruction with the diffusion prior (Section \ref{posterior_sampling}).

\subsection{Diffusion models}
\label{prior_sampling}
\begin{figure}[t!]
  \centering
  \begin{tikzpicture}
    \foreach \idx [count=\i from 0] in {0,4,3,6,2,5,1}{
      \node[inner sep=0pt] (img\i) at (\i*0.13\textwidth, 0)
        {\includegraphics[width=0.15\textwidth]{Figures/dataset/test_sample_\idx.pdf}};
    }
    \node[inner sep=0pt, anchor=west] (cbar) at ([xshift=3mm]img6.east)
      {\includegraphics[height=0.39\textwidth]{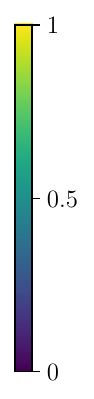}};
  \end{tikzpicture}
\captionsetup{justification=justified,singlelinecheck=false}
\caption{Training dataset samples: soft X-ray synthetic emissivity profiles.}
  \label{fig:training_data}
\end{figure}

Diffusion models \cite{song_2019, ho_2020, song_2021} instantiate the scheme in Eq.\eqref{annealed_ula} with a specific smoothing sequence: $p_t$ is the prior $p$ convolved with a Gaussian kernel of variance $\sigma_t^2$, where the noise levels $\sigma_t$ increase with $t$.
Equivalently, $p_t$ is the density of the noisy profile
\begin{equation}\label{diffusion_corruption}
    \mathbf{x}_t = \mathbf{x}_0 + \sigma_t\,\mathbf{z}\;,\qquad \mathbf{x}_0\sim p\;,\quad \mathbf{z}\sim\mathcal{N}(\mathbf{0},\mathbf{I})\;,
\end{equation}
so that the sequence $(p_t)$ describes the progressive corruption of a clean profile into pure noise.
This choice relates the scores required by Eq.\eqref{annealed_ula} to a denoising problem: by \emph{Tweedie's formula} \cite{efron2011}, the score of the noisy distribution $p_t$ is given by
\begin{equation}\label{tweedie_score}
    \nabla \log p_t(\mathbf{x}_t) = \frac{\mathbb{E}[\mathbf{x}_0\,\vert\,\mathbf{x}_t] - \mathbf{x}_t}{\sigma_t^2}\;,
\end{equation}
where the posterior mean $\mathbb{E}[\mathbf{x}_0\,\vert\,\mathbf{x}_t]$ is the Minimum Mean Square Error (MMSE) estimate (see, e.g., \cite[Sec.~5.1.2]{kaipio_statistical_2005}) of the clean profile $\mathbf{x}_0$ given its noisy version $\mathbf{x}_t$, i.e.,
\begin{equation}\label{mmse_estimate_}
\mathbb{E}[\mathbf{x}_0\,\vert\,\mathbf{x}_t]\!=\!\argmin_{\tilde{\mathbf{x}}\in\mathbb{R}^N}\int_{\mathbb{R}^N}\Vert\tilde{\mathbf{x}}-\mathbf{x}_0\Vert_2^2 \,p(\mathbf{x}_0\vert\mathbf{x}_t)\,\mathrm{d}\mathbf{x}_0\;\;.
\end{equation}
The right-hand side of Eq.\eqref{tweedie_score} represents the negative posterior-expected noise, rescaled by the noise level.
Learning the scores of the family $(p_t)$ is therefore equivalent to learning a Gaussian \emph{denoiser} for all noise levels $\sigma_t$.
\\
Whereas Eq.\eqref{mmse_estimate_} characterizes the estimate pointwise, for one noisy
observation $\mathbf{x}_t$ at a time, averaging the same quadratic loss over the joint
distribution of $(\mathbf{x}_0,\mathbf{x}_t)$ characterizes the denoiser as a whole.
For a candidate denoiser $f:\mathbb{R}^N\to\mathbb{R}^N$, writing
$p(\mathbf{x}_0,\mathbf{x}_t)=p(\mathbf{x}_0\vert\mathbf{x}_t)\,p_t(\mathbf{x}_t)$ gives
\begin{equation}\label{eq_loss_}
\begin{aligned}
\mathbb{E}_{(\mathbf{x}_0,\mathbf{x}_t)}\big[\Vert f(\mathbf{x}_t)-\mathbf{x}_0\Vert_2^2\big]
&=\int_{\mathbb{R}^N}\int_{\mathbb{R}^N}\Vert f(\mathbf{x}_t)-\mathbf{x}_0\Vert_2^2\,
p(\mathbf{x}_0,\mathbf{x}_t)\,\mathrm{d}\mathbf{x}_0\,\mathrm{d}\mathbf{x}_t\\
&=\int_{\mathbb{R}^N}\bigg[\int_{\mathbb{R}^N}\Vert f(\mathbf{x}_t)-\mathbf{x}_0\Vert_2^2\,
p(\mathbf{x}_0\vert\mathbf{x}_t)\,\mathrm{d}\mathbf{x}_0\bigg]\,p_t(\mathbf{x}_t)\,\mathrm{d}\mathbf{x}_t\,.
\end{aligned}
\end{equation}
The inner integral in the second line of Eq.\eqref{eq_loss_} is the conditional loss of Eq.\eqref{mmse_estimate_} evaluated at
$\tilde{\mathbf{x}}=f(\mathbf{x}_t)$.
Since $p_t\geq 0$ and since the posterior mean $\mathbb{E}[\mathbf{x}_0\vert\mathbf{x}_t]$ minimizes the inner integral for every $\mathbf{x}_t$, $f(\mathbf{x}_t)=\mathbb{E}[\mathbf{x}_0\vert\mathbf{x}_t]$ also minimizes the joint expectation at the left-hand side of Eq.\eqref{eq_loss_}.
The MMSE denoiser, i.e., the function of $\mathbf{x}_t$ that best predicts $\mathbf{x}_0$ in
the mean-square sense, is thus the posterior mean:\footnote{More precisely, the minimization runs over measurable functions $f$ that are square-integrable with respect to $p_t$, and the minimizer is unique up to $p_t$-null sets.}
\begin{equation}\label{denoising_loss}
\mathbb{E}[\mathbf{x}_0\,\vert\,\mathbf{x}_t=\cdot\,]=\argmin_{f:\mathbb{R}^N\to\mathbb{R}^N}
\mathbb{E}_{(\mathbf{x}_0,\mathbf{x}_t)}\big[\Vert f(\mathbf{x}_t)-\mathbf{x}_0\Vert_2^2\big]\,.
\end{equation}
For a fixed noise level $\sigma_t$, the denoiser in Eq.\eqref{denoising_loss} can be learned by least-squares regression on pairs $(\mathbf{x}_0, \mathbf{x}_t)$.
Crucially, the objective involves the unknown distributions $p$ and $p_t$ only through samples of pairs $(\mathbf{x}_0, \mathbf{x}_t)$. The expectation is approximated with a Monte Carlo estimate: clean profiles $\mathbf{x}_0$ are drawn from a dataset, and their noisy versions $\mathbf{x}_t$ are constructed as in Eq.\eqref{diffusion_corruption}.
Turning Eq.\eqref{denoising_loss} into a practical training scheme then requires two ingredients: (i) a class of functions over which the minimization performs well, the standard choice being a U-Net convolutional architecture \cite{ho_2020}; and (ii) a dataset of representative samples from the distribution of interest.
Training a denoiser on the resulting empirical loss thus yields an approximation of the MMSE-optimal denoiser $\mathbb{E}[\mathbf{x}_0\,\vert\,\mathbf{x}_t]$, and hence, through Tweedie's formula~\eqref{tweedie_score}, of the scores required by Eq.\eqref{annealed_ula}.\\
In practice, the network is often parametrized to predict the noise $\mathbf{z}$ rather than the clean profile; we denote this parametrization $\boldsymbol{\epsilon}_\theta(\mathbf{x}_t, t)$, with $\theta$ the trainable weights. By Eq.\eqref{diffusion_corruption}, the two parametrizations are equivalent, as $\mathbf{x}_0$ can be computed from $\mathbf{x}_t$ by removing the noise $\sigma_t\mathbf{z}$. Usually, a single network is trained jointly across all noise levels: the objective of Eq.\eqref{denoising_loss} is further averaged over the diffusion time $t$, and the network receives $t$ (equivalently, the noise level $\sigma_t$) as an additional input, so that one model implements the entire family of denoisers.
\\
For our application, we generate a training set of $10^5$ synthetic emissivity profiles with the procedure described in \cite{hamm2025}.
Each phantom is constructed from a realistic TCV magnetic equilibrium $\psi$, by summing a poloidal perturbation to a background emission profile.
The resulting phantoms include physically realistic Gaussian-shaped, peaked, and hollow profiles, as well as poloidal asymmetries of various kinds (in/out, top/bottom, and intermediate cases). Samples from the dataset are shown in Fig.\ref{fig:training_data}.
\\
A common choice for the noise schedule and the weighting of the training loss is the one of DDPM \cite{ho_2020}, in which the noisy profiles are constructed as $\mathbf{x}_t = \sqrt{\bar{\alpha}_t}\,\mathbf{x}_0 + \sqrt{1-\bar{\alpha}_t}\,\mathbf{z}$, with $\bar{\alpha}_t = \prod_{s\le t}(1-\beta_s)$ for a prescribed schedule of noise rates $\{\beta_t\}_{t=1}^{T}\subset(0,1)$.

\begin{figure}[t!]
  \raggedright
  \begin{tikzpicture}
    \foreach \step/\idx [count=\i from 0] in {
      1_0/0,
      151_1/1,
      211_2/2,
      235_3/3,
      244_4/4,
      246_5/5,
      249_6/6%
    }{
      \node[inner sep=0pt] (img\idx) at (\i*0.13\textwidth, 0)
        {\includegraphics[width=0.15\textwidth]{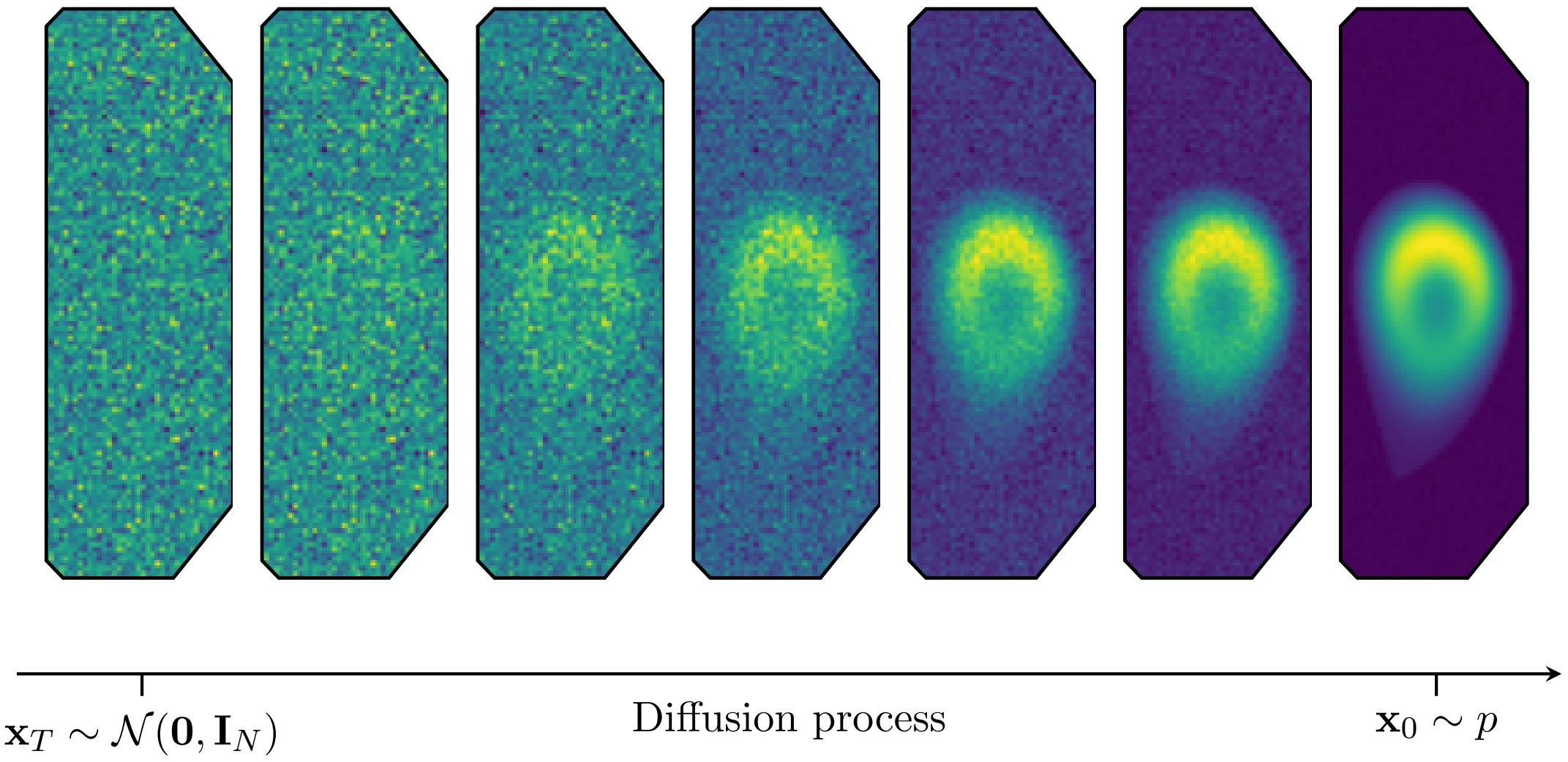}};
    }
    \draw[->, thick, >=stealth]
      ([yshift=-6mm]img0.south -| img0.west) -- ([yshift=-6mm]img6.south -| img6.east)
      node[midway, below=1mm] {\small Diffusion process};
    \draw[thick] ([yshift=-6mm]img0.south) -- ++(0,-2mm)
      node[below, align=center] {\small $\mathbf{x}_T \sim \mathcal{N}(\mathbf{0}, \mathbf{I}_N)$};
    \draw[thick] ([yshift=-6mm]img6.south) -- ++(0,-2mm)
      node[below, align=center] {\small $\mathbf{x}_0 \sim p$};
  \end{tikzpicture}
\captionsetup{justification=justified,singlelinecheck=false}
  \caption{Prior sampling process: generation of a sample $\mathbf{x}_0$ of the learned prior $p$ from an initial noise realization $\mathbf{x}_T$.}  \label{fig:prior_sampling_process}
\end{figure}

\begin{figure}[t!]
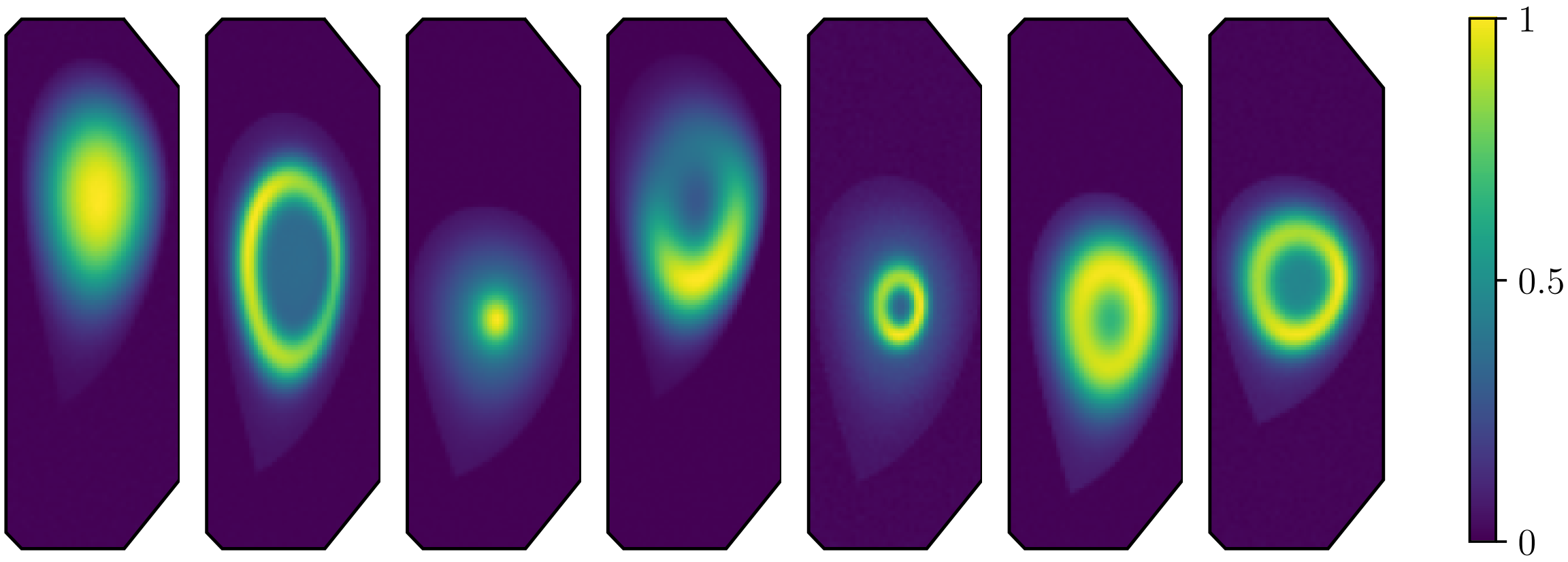

  \centering
  \begin{tikzpicture}
    \foreach \idx [count=\i from 0] in {0,4,3,5,2,6,1}{
      \node[inner sep=0pt] (img\i) at (\i*0.13\textwidth, 0)
        {\includegraphics[width=0.15\textwidth]{Figures/prior_samples/prior_sample_\idx.pdf}};
    }
    \node[inner sep=0pt, anchor=west] (cbar) at ([xshift=3mm]img6.east)
      {\includegraphics[height=0.39\textwidth]{Figures/dataset/phantoms_cbar.pdf}};
  \end{tikzpicture}
  \captionsetup{justification=justified,singlelinecheck=false}
  \caption{Prior samples, drawn from the learned prior using the DDPM algorithm.}
  \label{fig:prior_samples}
\end{figure}

This construction matches Eq.\eqref{diffusion_corruption} up to the scaling factor $\sqrt{\bar{\alpha}_t}$, with the effective noise level $\bar{\sigma}_t = \sqrt{(1-\bar{\alpha}_t)/\bar{\alpha}_t}$.
The DDPM schedule is said to be variance-preserving. This name reflects the continuous-time picture of \cite{song_2021}: the DDPM corruption discretizes the Ornstein--Uhlenbeck process
\begin{equation}\label{vp_sde}
    \mathrm{d}\mathbf{x} = -\tfrac{1}{2}\beta(t)\,\mathbf{x}\,\mathrm{d}t + \sqrt{\beta(t)}\,\mathrm{d}\mathbf{w}\;,
\end{equation}
where $\mathbf{w}$ is the standard Brownian motion\footnote{Eq.\eqref{vp_sde} is a stochastic differential equation (SDE): we refer the reader to our discussion in \cite{hamm2025} for a brief description of SDEs, Brownian motions, and stationary distributions.}.
The Ornstein--Uhlenbeck process converges to its stationary distribution $\mathcal{N}(\mathbf{0},\mathbf{I})$ regardless of the initial distribution; in this sense, the total variance remains bounded, hence the variance-preserving label.
Its time reversal is again a diffusion process \cite{anderson_1982, song_2021},
\begin{equation}\label{reverse_sde}
    \mathrm{d}\mathbf{x} = \Big[-\tfrac{1}{2}\beta(t)\,\mathbf{x} - \beta(t)\,\nabla\log p_t(\mathbf{x})\Big]\,\mathrm{d}t + \sqrt{\beta(t)}\,\mathrm{d}\bar{\mathbf{w}}\;,
\end{equation}
run backward from $t=T$ to $t=0$, where $\bar{\mathbf{w}}$ is the time-reversed standard Brownian motion.
In Eq.\eqref{reverse_sde}, the score acts as an extra drift term steering toward the data distribution.
The distribution $\pi_t$ that characterizes the process at time $t$ evolves from a prescribed $\pi_T$ (that approximates the true, unknown $p_T$), to a distribution $\pi_0$ (that hence approximates the prior $p$).
\\
In practice, DDPM sampling from the prior proceeds by \emph{iterative} denoising.
Generation starts from pure noise, $\mathbf{x}_T\sim\mathcal{N}(\mathbf{0},\mathbf{I})$, and applies $T$ stochastic transitions $\mathbf{x}_t\to\mathbf{x}_{t-1}$, each removing a small fraction of the noise.
One may ask why the learned denoiser is not simply applied once to $\mathbf{x}_T$.
The reason is that the network approximates the posterior \emph{mean} $\mathbb{E}[\mathbf{x}_0\,\vert\,\mathbf{x}_t]$: at the largest noise levels $\mathbf{x}_t$ carries almost no information about $\mathbf{x}_0$, so this mean collapses toward the average of the training distribution which is a blurred superposition of all plausible profiles rather than a sample from $p$.
The iterative scheme avoids this collapse.
At each step, the denoiser provides the current MMSE estimate $\hat{\mathbf{x}}_0(\mathbf{x}_t)$; the transition moves $\mathbf{x}_t$ only partway toward this estimate and re-injects a small amount of fresh noise, according to a Gaussian transition kernel whose explicit form is given in Appendix~\ref{app:sampling}.
The injected randomness is what turns the regression into a generative process: the early, coarse steps commit the trajectory to one region of the distribution, and the estimate $\hat{\mathbf{x}}_0$ is recomputed and sharpens as the noise level decreases, so that later steps refine fine-scale structure.
Each transition only needs to undo a small noise increment, a much easier task than inverting the full corruption in one shot, and the scheme as a whole is a discretization of the reverse-time dynamics of Eq.\eqref{reverse_sde} \cite{ho_2020, song_2021}.
The full procedure is summarized in Algorithm~\ref{alg:ddpm_sampling}.\\
The learned denoiser is a small U-Net with a three-level encoder–decoder structure, with $\approx 2\cdot10^5$ trainable parameters.
To train it, noisy profiles are constructed at different noise levels, indexed by the diffusion time step $t\in\{1,\ldots,T\}$; information about $t$ is embedded into the network, making its predictions dependent on diffusion time and thus on the noise level.
As customary in U-Nets, skip connections concatenate the features of each encoder level with those of the corresponding decoder level, preventing the loss of fine details; for a full description of the architecture used, we refer the reader to the paper repository.
For both training and deployment, a single NVIDIA Tesla V100 (32 GB) GPU was used.
The model was trained for $\approx 16$ hours.\\
Fig.~\ref{fig:prior_sampling_process} illustrates the resulting sampling process, which progressively transports pure noise onto a sample from the learned prior.
Fig.~\ref{fig:prior_samples} shows samples generated by the trained model.
No magnetic equilibrium is overlaid on the samples, because it is not available: the prior is trained to predict the emissivity profiles only.
The obtained samples are visually indistinguishable from the ones shown in Fig.~\ref{fig:training_data}, indicating that the prior learned to generate profiles compatible with the training data.\\

\subsection{Posterior sampling with diffusion priors}
\label{posterior_sampling}
\begin{algorithm}[t!]
\caption{DiffPIR \cite{zhu_2023} (black) and DiffRD (\textcolor{blue}{blue}) posterior sampling}
\label{alg:diffrd}
\begin{algorithmic}[1]
\Require Forward model $\mathbf{T} \in \mathbb{R}^{M \times N}$ and measurement $\mathbf{y}$ as in Eq.~\eqref{ip_data}; measurement noise std $\eta$ as in Eq.~\eqref{gaussian_likelihood}; noise-prediction network $\boldsymbol{\epsilon}_\theta$ (Section~\ref{prior_sampling}); number of steps $T$ and schedule $\{\bar{\alpha}_t\}_{t=1}^{T}$, with $\bar{\sigma}_t = \sqrt{(1-\bar{\alpha}_t)/\bar{\alpha}_t}$; fidelity weight $\lambda_{\mathrm{pir}}$; stochasticity $\zeta \in [0,1]$;
\textcolor{blue}{clipping mask $\mathbf{M} \in \{0,1\}^N$; diffusion tensor $\mathbf{D}(\psi,\alpha)$ and discrete gradient $\boldsymbol{\nabla}$ as in Eq.~\eqref{eq_diffusion_prior}; RD strength $\lambda_\mathrm{rd}$}
\State Initialize $\mathbf{x}_T \sim \mathcal{N}(\mathbf{0}, \mathbf{I})$; pre-calculate $\rho_t \triangleq \lambda_{\mathrm{pir}}\,\eta^2 / \bar{\sigma}_t^2$.
\For{$t = T$ \textbf{to} $1$}
    \State $\hat{\mathbf{x}}_0^{(t)} = \frac{1}{\sqrt{\bar{\alpha}_t}}\bigl(\mathbf{x}_t - \sqrt{1 - \bar{\alpha}_t}\;\boldsymbol{\epsilon}_\theta(\mathbf{x}_t, t)\bigr)$
    \Statex \quad\textit{// MMSE prediction of the clean profile, as in Algorithm~\ref{alg:ddpm_sampling}}
    \State $\tilde{\mathbf{x}}_0^{(t)} = \arg\min_{\mathbf{x}} \|\mathbf{y} - \mathbf{T}\mathbf{x}\|^2 + \rho_t\|\mathbf{x} - \hat{\mathbf{x}}_0^{(t)}\|^2$
    \textcolor{blue}{${}+ \lambda_\mathrm{rd}\,\bar{\sigma}_t\,\|\sqrt{\mathbf{D}(\psi,\alpha)}\,\boldsymbol{\nabla}\mathbf{x}\|^2$}
    \Statex \quad\textit{// Proximal data subproblem} \textcolor{blue}{\textit{with RD regularization (ours)}}
    \State \textcolor{blue}{$\tilde{\mathbf{x}}_0^{(t)} \leftarrow \tilde{\mathbf{x}}_0^{(t)} \odot \mathbf{M}$}
    \Statex \quad\textcolor{blue}{\textit{// Geometry clipping (ours)}}
    \State $\hat{\boldsymbol{\epsilon}} = \frac{1}{\sqrt{1-\bar{\alpha}_t}}\bigl(\mathbf{x}_t - \sqrt{\bar{\alpha}_t}\,\tilde{\mathbf{x}}_0^{(t)}\bigr)$
    \Statex \quad\textit{// Effective noise consistent with the corrected estimate}
    \State $\boldsymbol{\epsilon}_t \sim \mathcal{N}(\mathbf{0}, \mathbf{I})$
    \State $\mathbf{x}_{t-1} = \sqrt{\bar{\alpha}_{t-1}}\,\tilde{\mathbf{x}}_0^{(t)} + \sqrt{1-\bar{\alpha}_{t-1}}\bigl(\sqrt{1-\zeta}\;\hat{\boldsymbol{\epsilon}} + \sqrt{\zeta}\;\boldsymbol{\epsilon}_t\bigr)$
    \Statex \quad\textit{// One likelihood-corrected reverse step}
\EndFor
\State \Return $\mathbf{x}_0$
\end{algorithmic}
\end{algorithm}

\begin{figure}[t!]
  \raggedright
  \begin{tikzpicture}
    \node[inner sep=0pt] (gt) at (0, 0.4\textwidth)
      {\includegraphics[width=0.15\textwidth]{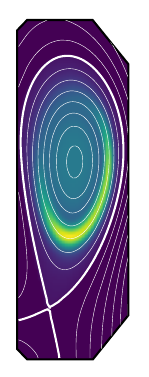}};
    \node[anchor=south, font=\normalsize] at ([yshift=-2mm]gt.north) {$\mathbf{x}$};
    \node[rotate=90, anchor=south, font=\small] at ([xshift=-4mm]gt.west) {inverse$\,$problem$:\,\mathbf{T}\mathbf{x}=\mathbf{y}$};
    \node[inner sep=0pt, anchor=west] (cbar) at ([xshift=-2mm]gt.east)
      {\includegraphics[height=0.39\textwidth]{Figures/dataset/phantoms_cbar.pdf}};
    \node[inner sep=0pt] (los) at (0.26\textwidth, 0.4\textwidth)
      {\includegraphics[width=0.15\textwidth]{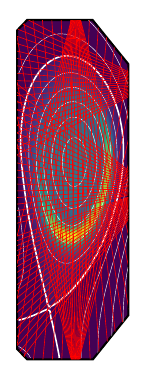}};
    \node[anchor=south, font=\normalsize] at ([yshift=-2mm]los.north) {$\mathbf{T}$};
    \node[inner sep=0pt, anchor=west] (tomo) at ([xshift=0.07\textwidth, yshift=2.5mm]los.east) {%
      \begin{tikzpicture}
        \begin{axis}[
      width=0.48\textwidth,
      height=0.39\textwidth,
      xmin=-2, xmax=103,
      xtick={1,20,40,60,80,100},
      xticklabels={$1$,$20$,$40$,$60$,$80$,$100$},
      xtick align=outside,
      xtick pos=bottom,
      minor tick num=0,
      tick style={thin, major tick length=3pt},
      xlabel={$\mathrm{LoS\,index}$},
      xlabel style={font=\small, yshift=-4pt},
      tick label style={font=\small},
      ytick align=outside,
      ytick pos=left,
      ytick={0, 10},
      yticklabels={$0$, $10$},
      enlarge y limits={value=0.08, auto},
      axis line style={-},
      title={Tomographic data},
      title style={font=\normalsize, yshift=-1pt},
      legend style={
        at={(0.05,0.98)},
        anchor=north west,
        font=\small,
        fill=white,
        fill opacity=1,
        draw opacity=1,
        text opacity=1,
        legend columns=1,
      },
      clip=false,
    ]
      \addplot[
        only marks,
        mark=*,
        mark size=0.8pt,
        color={rgb,255:red,31;green,119;blue,180},
      ] table[
        x expr=\coordindex+1, y=y_true, col sep=comma,
      ] {Figures/posterior_sampling_process/sxr_tomo_data.csv};
      \addlegendentry{\hspace{0.5em}$\mathbf{y}_{\mathrm{true}}$}
      \addplot[
        only marks,
        mark=x,
        mark size=1.5pt,
        mark options={line width=0.8pt},
        red,
      ] table[
        x expr=\coordindex+1, y=y_noisy, col sep=comma,
      ] {Figures/posterior_sampling_process/sxr_tomo_data.csv};
      \addlegendentry{\hspace{-1em}$\mathbf{y}$}
    \end{axis}
      \end{tikzpicture}%
    };
    \foreach \step/\idx [count=\i from 0] in {
      0_0/0,
      35_1/1,
      71_2/2,
      106_3/3,
      142_4/4,
      177_5/5,
      249_6/6%
    }{
      \node[inner sep=0pt] (dif\idx) at (\i*0.13\textwidth, 0)
        {\includegraphics[width=0.15\textwidth]{Figures/posterior_sampling_process/diffusion_process_sxr_step_\step.pdf}};
    }
    \foreach \step/\idx [count=\i from 0] in {
      0_0/0,
      35_1/1,
      71_2/2,
      106_3/3,
      177_4/4,
      213_5/5,
      249_6/6%
    }{
      \node[inner sep=0pt] (drd\idx) at (\i*0.13\textwidth, -0.38\textwidth)
        {\includegraphics[width=0.15\textwidth]{Figures/posterior_sampling_process/diffrd_process_sxr_step_\step.pdf}};
    }
    \node[rotate=90, anchor=south, font=\small] at ([xshift=-4mm]dif0.west) {Diff$\,$PIR};
    \node[rotate=90, anchor=south, font=\small] at ([xshift=-4mm]drd0.west) {Diff$\,$RD};
    \draw[->, thick, >=stealth]
      ([yshift=-6mm]drd0.south -| drd0.west) -- ([yshift=-6mm]drd6.south -| drd6.east)
      node[midway, below=1mm] {\small Diffusion process};
    \draw[thick] ([yshift=-6mm]drd0.south) -- ++(0,-2mm)
      node[below, align=center] {\small $\widehat{\mathbf{x}}_0^{(t=T)}$};
    \draw[thick] ([yshift=-6mm]drd6.south) -- ++(0,-2mm)
      node[below, align=center] {\small $\widehat{\mathbf{x}}_0^{(t=0)}$};
  \end{tikzpicture}
  \captionsetup{justification=justified,singlelinecheck=false}
  \caption{Posterior sampling process: SXR tomographic reconstruction. Top: visual representation of the inverse problem of estimating the ground truth $\mathbf{x}$ from the noisy data $\mathbf{y}$ acquired with SXR diagnostic. Middle and bottom: generation of one posterior sample with DiffPIR and DiffRD, respectively; the shown profiles evolve from the MMSE prediction $\widehat{\mathbf{x}}_0^{(t=T)}$ at time $t=T$ to the prediction $\widehat{\mathbf{x}}_0^{(t=0)}$ of the final step, which corresponds to a clean posterior sample.}
  \label{fig:posterior_sampling_process}
\end{figure}

The reconstruction task requires samples from the posterior $p(\mathbf{x}\vert\mathbf{y})$ rather than from the prior.
Our framework retains the Bayesian decomposition of Eq.~\eqref{bayes_formula}, which disentangles the prior from the likelihood: the prior is learned once, while the likelihood encodes the diagnostic at hand.
The same learned prior thus serves any forward model without retraining.
Running the reverse diffusion of Eq.~\eqref{reverse_sde} on the posterior requires the scores of its smoothed versions, which Bayes' rule splits into
\begin{equation}\label{posterior_score_split}
    \nabla \log p_t(\mathbf{x}_t\vert\mathbf{y}) = \nabla\log p_t(\mathbf{x}_t) + \nabla\log p_t(\mathbf{y}\vert\mathbf{x}_t)\;.
\end{equation}
The first term is the prior score learned in Section~\ref{prior_sampling}.
The second term is intractable: the noise model in Eq.~\eqref{gaussian_likelihood} specifies the likelihood $p(\mathbf{y}\vert\mathbf{x}_0)$, but the evaluation of $p_t(\mathbf{y}\vert\mathbf{x}_t)=\int p(\mathbf{y}\vert\mathbf{x}_0)p(\mathbf{x}_0\vert\mathbf{x}_t)\,\mathrm{d}\mathbf{x}_0$ requires marginalizing $p(\mathbf{y}\vert\mathbf{x}_0)$ over all clean profiles compatible with $\mathbf{x}_t$, i.e., over the denoising posterior $p(\mathbf{x}_0\vert\mathbf{x}_t)$.
\\
Sampling algorithms differ in how they approximate this term. As discussed in Section~\ref{introduction}, popular examples include DPS \cite{chung_2024} and DiffPIR \cite{zhu_2023}.
In this work, we adopt DiffPIR, motivated by its strong performance in the statistical benchmark of \cite{zach_2025}.
DiffPIR is inspired by half-quadratic splitting~\cite{hqs}, an optimization technique that alternates between a prior subproblem and a likelihood subproblem.
It retains the reverse-time dynamics of Section~\ref{prior_sampling}, replacing each transition $\mathbf{x}_t\to\mathbf{x}_{t-1}$ with a likelihood-corrected one; the procedure is reported in Algorithm~\ref{alg:diffrd} (black lines, reproduced with minor notational adaptations from \cite{zhu_2023}).
At each step $t$, the network $\boldsymbol{\epsilon}_\theta$ first denoises the current iterate into the MMSE prediction $\hat{\mathbf{x}}_0^{(t)}$ of the clean profile, exactly as in the prior sampler of Algorithm~\ref{alg:ddpm_sampling} (prior step).
A proximal subproblem then balances fidelity to the measurements against proximity to this prediction (likelihood step), with a coupling $\rho_t = \lambda_{\mathrm{pir}}\,\eta^2/\bar{\sigma}_t^2$ that strengthens as the effective noise level $\bar{\sigma}_t$ decreases.
Since $\bar{\sigma}_t$ decays from its maximal value at $t=T$ (pure noise) to zero at $t=0$ (clean sample), early iterations trust the measurements while later ones increasingly trust the denoiser, whose prediction sharpens as the noise decreases.
The corrected estimate $\tilde{\mathbf{x}}_0^{(t)}$ is finally mapped back to the noise level of step $t-1$: the effective noise $\hat{\boldsymbol{\epsilon}}$ is mixed with fresh Gaussian noise in proportions set by $\zeta\in[0,1]$, ranging from a deterministic update ($\zeta=0$) to full re-noising ($\zeta=1$).
For $\zeta=0$, the whole generative process is deterministic given the initial draw $\mathbf{x}_T$, which is the only source of variability\footnote{For the interested reader: if the likelihood step from DiffPIR is removed, the $\zeta=0$ regime yields the deterministic DDIM updates \cite{song2021denoising}, whose underlying time-continuous interpretation is given by the probability flow ODE of \cite{song_2021}.}.
For $\zeta=1$, the effective noise $\hat{\boldsymbol{\epsilon}}$ is discarded entirely, and $\tilde{\mathbf{x}}_0^{(t)}$ is re-noised from scratch to level $t-1$ through the forward DDPM construction introduced in Section \ref{prior_sampling}, with $\mathbf{x}_{t-1} = \sqrt{\bar{\alpha}_{t-1}}\,\tilde{\mathbf{x}}_0^{(t)} + \sqrt{1-\bar{\alpha}_{t-1}}\,\mathbf{z}$, $\mathbf{z}\sim\mathcal{N}(\mathbf{0}, \mathbf{I})$; variability across runs is maximal.
The iteration proceeds until $t=0$, and each run of the algorithm returns one approximate posterior sample.
As runs are independent, samples can be drawn simultaneously as a batch on modern GPUs; the resulting independent posterior samples can then be used to compute Monte Carlo estimates of quantities of interest, e.g., the posterior mean $\vmu$ and standard deviation $\vsigma$.
\\
In addition, we introduce DiffRD, an adaptation of DiffPIR to the cases in which the magnetic equilibrium is known; the required modifications are reported in blue in Algorithm~\ref{alg:diffrd}.
DiffRD augments the proximal subproblem with the equilibrium-informed RD regularization of Eq.~\eqref{eq_diffusion_prior}, with a strength $\lambda_{\mathrm{rd}}\,\bar{\sigma}_t$ that decays along the reverse process,\footnote{The weight $\lambda_{\mathrm{rd}}$ is a hyperparameter of the sampler: it plays a role analogous to, but is distinct from and tuned independently of, the regularization weight $\lambda$ of Eq.~\eqref{eq_diffusion_prior}.} and restricts the estimate to the reconstruction region through the clipping mask $\mathbf{M}$, enforcing zero emissivity outside the core.
The modification thus enables posterior sampling with reaction--diffusion guidance while preserving the structure and the cost of DiffPIR: effectively, DiffRD samples under a hybrid prior, combining the data-driven diffusion prior with the hand-designed, equilibrium-informed RD one.
\\
In our experiments, we set the number of diffusion steps of DiffPIR and DiffRD to $T=250$; for large enough $T$, performance is not very sensitive to its choice.
Once $T$ is set, the scaling factors $\{\bar{\alpha}_t\}_{t=1}^T$ and effective noise levels $\{\bar{\sigma}_t\}_{t=1}^T$ are known.
The only hyperparameters to be tuned for a given application are $\lambda_{\mathrm{pir}}$ and $\zeta$, plus $\lambda_{\mathrm{rd}}$ (and potentially the anisotropy parameter $\alpha$) for DiffRD.
\\
Figure~\ref{fig:posterior_sampling_process} illustrates the resulting posterior sampling process, and makes the interplay between the two terms of Eq.~\eqref{posterior_score_split} visible.
In DiffPIR, the early iterations are dominated by data fidelity: line-of-sight streaks appear in the iterates, the imprint of the likelihood pulling the profile toward the measured chords.
As $\bar{\sigma}_t$ decreases, the balance shifts toward the prior, and the final steps essentially denoise the coarse, measurement-consistent guess assembled earlier.
In DiffRD, the RD term, strongest at early times through the factor $\lambda_{\mathrm{rd}}\,\bar{\sigma}_t$, drives the iterate within a few steps toward the smooth, equilibrium-aligned RD solution; the weight then progressively shifts to the learned denoiser, which dominates the late iterations and restores fine-scale structure.
We compare the RD, DiffPIR, and DiffRD frameworks in Section~\ref{results}.

\clearpage

\section{Results: Phantoms and Experimental Data}
\label{results}
In this section, we first discuss the results of our phantom studies (\ref{sxr_vs_dmpx}, \ref{pilatus}), then consider the application to TCV experimental data (\ref{experimental_data}), and finally examine potential limitations of the diffusion framework (\ref{limitations}).\\
Before presenting the results, we describe the protocol common to the phantom studies of Sections \ref{sxr_vs_dmpx} (SXR and DMPX) and \ref{pilatus} (Pilatus).
We evaluate the methods on a test set of $1000$ synthetic phantoms, generated with the procedure of Section \ref{prior_sampling} and held out from training.
For each phantom and each diagnostic, we simulate the tomographic measurements according to Eq.\eqref{ip_data}, using the geometry matrices introduced in Section \ref{inverse_problems_tomography} and the Gaussian noise model of Eq.\eqref{gaussian_likelihood}.
We consider three noise levels per diagnostic, denoted by $\eta_{\text{low}} < \eta_{\text{medium}} < \eta_{\text{high}}$.
Similarly to our previous studies \cite{hamm2025}, these signal- and detector-independent noise levels correspond to approximately $1\%$, $5\%$ and $10\%$ of the average tomographic measurement of the diagnostic computed over the dataset, respectively.
For reference, the noisy data $\mathbf{y}$ shown in Fig.\ref{fig:posterior_sampling_process} for the SXR diagnostic are obtained for $\eta=\eta_{\text{high}}$.\\
All the considered methods (RD, DiffPIR and DiffRD) have hyperparameters that need to be defined or tuned at reconstruction time.
For all methods and diagnostics, we assume that the noise level $\eta$ is known.
For RD, the tuning strategy adopted in \cite{hamm2025} was found to be satisfactory for all diagnostics and noise levels.
We thus use a regularization parameter $\lambda=10^{-1}$, and tune the anisotropic parameter $\alpha$ individually for each reconstruction, with a
procedure that selects the value that best fits a subset of noisy tomographic data kept hidden from the reconstruction algorithm.
For DiffPIR and DiffRD, a grid search on a few representative samples shows that the hyperparameters $\lambda_{pir}$ and $\zeta$ can be set once per diagnostic and kept fixed across the whole test set.
The same holds for the hyperparameter $\lambda_{rd}$ in DiffRD.
The selected values will be reported in Sections \ref{sxr_vs_dmpx} and \ref{pilatus}.
No reconstruction-specific tuning is thus needed for DiffPIR, while for DiffRD $\alpha$ is tuned with the same procedure implemented for RD.
\\
For each method, we estimate the posterior mean $\vmu$ and the pixel-wise posterior standard deviation $\vsigma$; we take $\vmu$ as our reconstruction, and quantify its uncertainty with $\vsigma$.
They are computed as the componentwise Monte Carlo estimates
\begin{equation}\label{posterior_mc_estimates}
    \mu_n \approx \frac{1}{S}\sum_{s=1}^{S} x^{(s)}_n\;,\qquad
    \sigma_n \approx \sqrt{\frac{1}{S-1}\sum_{s=1}^{S}\big(x^{(s)}_n - \mu_n\big)^2}\;,\qquad n = 1,\ldots,N\;,
\end{equation}
where the subscript $n$ denotes the $n$-th pixel and $\mathbf{x}^{(1)},\ldots,\mathbf{x}^{(S)}$ are the drawn posterior samples.
For the RD baseline, as in \cite{hamm2025}, we rely on ULA, discarding the first $10^3$ samples in a burn-in phase and then computing the estimates using the following $10^5$ samples; as discussed in \cite{hamm2025}$, \vmu$ and $\vsigma$ are updated by efficient online updates after each draw, so that it is not necessary to store all the samples as Eq.\eqref{posterior_mc_estimates} would suggest.
For the diffusion framework, we draw $100$ samples with DiffPIR/DiffRD, then evaluate $\vmu$ and $\vsigma$ by Eq.\eqref{posterior_mc_estimates}.
While each run of DiffPIR/DiffRD yields independent posterior samples, ULA produces a Markov chain whose successive samples are correlated between each other, which is why it requires more samples to deliver accurate estimates.
Moreover, DiffPIR/DiffRD samples can be drawn in parallel, whereas a single ULA chain is inherently sequential.
In turn, however, the per-sample cost of ULA is much lower than DiffPIR/DiffRD.
Ultimately, each full ULA run took $\approx50\,\mathrm{s}$ on a workstation with Core i9-10900X processor, while drawing 100 samples in parallel took $\approx10\,\mathrm{s}$ for DiffPIR and $\approx14\,\mathrm{s}$ for DiffRD on the hardware specified in Section \ref{prior_sampling}.
\\
We measure the quality of the posterior mean $\vmu$ against the ground truth $\mathbf{x}_{_{GT}}$ with the Peak Signal-to-Noise Ratio (PSNR, in decibels) and the Structural Similarity Index Measure (SSIM) \cite{Wang2004}.
The first is defined as $\mathrm{PSNR}=10\log_{10}(d_r^2/\mathrm{MSE})$, where MSE is the mean squared error between $\vmu$ and $\mathbf{x}_{_{GT}}$, and $d_r=(\max_n x_{_{GT,n}}-\min_n x_{_{GT,n}})$ is the dynamic range of the ground truth.
The dependence on $d_r$
makes the computed PSNR values comparable across phantoms of differing range.
SSIM, instead, measures the perceived visual quality.
Higher is better for both metrics.
In the tables, we report the mean and the standard deviation of each metric over the test set, we mark the best value in bold, and we star it when the best method is significantly better than each competitor according to a one-sided paired Wilcoxon signed-rank test at $p<0.05$ \cite{demsvar_2006}.
Unlike a comparison in terms of average performance on the dataset, which can be dominated by a few phantoms, the signed-rank test asks whether the per-phantom differences are consistently in favor of a given method.

\subsection{Phantom study: SXR and DMPX diagnostics}
\label{sxr_vs_dmpx}
\begin{figure}[t!]
\centering
\newcommand{\panelW}{0.15\textwidth}
\newcommand{\panelH}{0.39\textwidth}
\newcommand{\panelSep}{0.04cm}
\newcommand{\groupSep}{0.55cm}
\newcommand{\cbarSep}{0.08cm}
\newcommand{\cbarW}{0.30cm}
\newcommand{\rowSep}{1.0cm}
\newcommand{\zoomCx}{0.50}
\newcommand{\zoomCy}{0.42}
\newcommand{\zoomHw}{0.15}
\newcommand{\zoomHh}{0.08}
\newcommand{\zoomMag}{2.5}
\pgfmathsetlengthmacro{\srcW}{2 * \zoomHw * \panelW}
\pgfmathsetlengthmacro{\srcH}{2 * \zoomHh * \panelH}
\pgfmathsetlengthmacro{\outW}{\zoomMag * \srcW}
\pgfmathsetlengthmacro{\outH}{\zoomMag * \srcH}
\begin{tikzpicture}[
  panel/.style={anchor=south west, inner sep=0pt, outer sep=0pt},
  spy using outlines={
    rectangle,
    width=\outW,
    height=\outH,
    magnification=\zoomMag,
    connect spies,
    every spy on node/.style ={draw=red, very thick},
    every spy in node/.style ={draw=red, very thick},
  },
]
  \def\D{Figures/diffpir_vs_rd/sxr_sigma0.25_idx661}

  \node[panel] (phantom) at (0,0)
    {\includegraphics[width=\panelW]{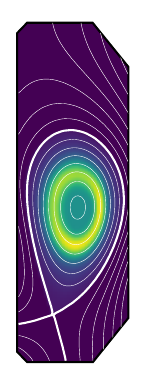}};
  \node[panel] (rdmean)  at ($(phantom.south east)+(\panelSep,0)$)
    {\includegraphics[width=\panelW]{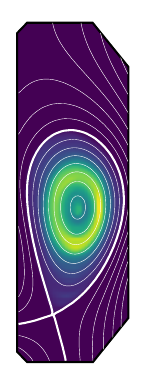}};
  \node[panel] (dpmean)  at ($(rdmean.south east)+(\panelSep,0)$)
    {\includegraphics[width=\panelW]{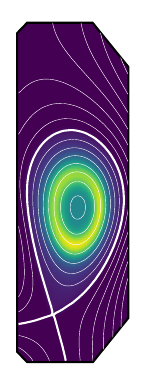}};
  \node[panel, anchor=west] (cbarM) at ($(dpmean.east)+(\cbarSep,0)$)
{\includegraphics[height=0.39\textwidth]{Figures/dataset/phantoms_cbar.pdf}};

  \node[panel] (rdstd)   at ($(cbarM.south east)+(\groupSep,0)$)
    {\includegraphics[width=\panelW]{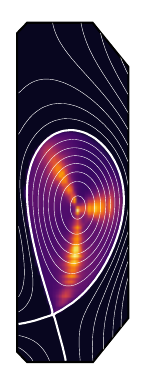}};
  \node[panel] (cbarS1)  at ($(rdstd.south east)+(\cbarSep,0)$)
    {\includegraphics[height=0.39\textwidth]{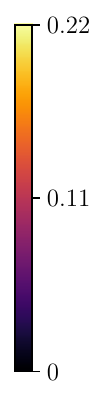}};
  \node[panel, anchor=west] (dpstd)   at ($(cbarS1.east)+(\panelSep,0)$)
    {\includegraphics[width=\panelW]{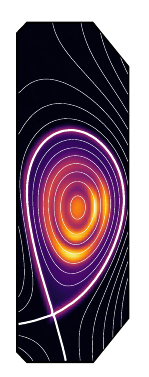}};
  \node[panel, anchor=west] (cbarS2)  at ($(dpstd.east)+(\cbarSep,0)$)
    {\includegraphics[height=0.39\textwidth]{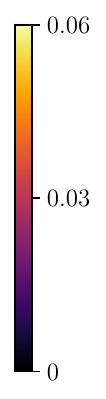}};

\node[anchor=south, font=\small] at ($(phantom.north) + (0, -0.3)$) {$\vx_{_\text{GT}}$};
\node[anchor=south, font=\small] at ($(rdmean.north) + (0, -0.3)$)  {$\vmu_{_\text{RD}}$};
\node[anchor=south, font=\small] at ($(dpmean.north) + (0, -0.3)$)  {$\vmu_{_\text{DiffPIR}}$};
\node[anchor=south, font=\small] at ($(rdstd.north) + (0, -0.3)$)   {$\vsigma_{_\text{RD}}$};
\node[anchor=south, font=\small] at ($(dpstd.north) + (0, -0.3)$)   {$\vsigma_{_\text{DiffPIR}}$};
  \foreach \n in {phantom, rdmean, dpmean}{
    \coordinate (\n-zc) at
      ($(\n.south west)!\zoomCx!(\n.south east)
       !\zoomCy!
       ($(\n.north west)!\zoomCx!(\n.north east)$)$);
  }

  \coordinate (z-phantom) at ($(phantom.south)+(0,-\rowSep-0.25*\outH)$);
  \coordinate (z-rdmean)  at ($(rdmean.south) +(0,-\rowSep-0.25*\outH)$);
  \coordinate (z-dpmean)  at ($(dpmean.south) +(0,-\rowSep-0.25*\outH)$);

  \spy on (phantom-zc) in node at (z-phantom);
  \spy on (rdmean-zc)  in node at (z-rdmean);
  \spy on (dpmean-zc)  in node at (z-dpmean);

\node[anchor=north, font=\small]
  at ($(rdstd.south)!0.5!(dpstd.south) + (0, -\rowSep + 0.7cm)$)
  {%
    \renewcommand{\arraystretch}{1.4}%
    \begin{tabular}{@{}lcc@{}}
    \hline
                              & PSNR           & SSIM           \\
    \hline
    $\vmu_{_\text{RD}}$      & 33.11          & 0.966          \\
    $\vmu_{_\text{DiffPIR}}$ & \textbf{40.42} & \textbf{0.991} \\
    \hline
    \end{tabular}%
  };

\end{tikzpicture}
\captionsetup{justification=justified,singlelinecheck=false}
\caption{SXR diagnostic: reconstruction of the ground truth $\mathbf{x}_{_{GT}}$, for noise level $\eta=\eta_{medium}$. For both RD and DiffPIR, the posterior mean $\vmu$ and standard deviation $\vsigma$ are shown; the value of the performance metrics is also reported.}
\label{fig:diffpir_vs_rd_sxr}
\end{figure}

\begin{figure}[t!]
\centering
\newcommand{\panelW}{0.15\textwidth}
\newcommand{\panelH}{0.39\textwidth}
\newcommand{\panelSep}{0.04cm}
\newcommand{\groupSep}{0.55cm}
\newcommand{\cbarSep}{0.08cm}
\newcommand{\cbarW}{0.30cm}
\newcommand{\rowSep}{1.0cm}
\newcommand{\zoomCx}{0.50}
\newcommand{\zoomCy}{0.37}
\newcommand{\zoomHw}{0.15}
\newcommand{\zoomHh}{0.08}
\newcommand{\zoomMag}{2.5}
\pgfmathsetlengthmacro{\srcW}{2 * \zoomHw * \panelW}
\pgfmathsetlengthmacro{\srcH}{2 * \zoomHh * \panelH}
\pgfmathsetlengthmacro{\outW}{\zoomMag * \srcW}
\pgfmathsetlengthmacro{\outH}{\zoomMag * \srcH}
\begin{tikzpicture}[
  panel/.style={anchor=south west, inner sep=0pt, outer sep=0pt},
  spy using outlines={
    rectangle,
    width=\outW,
    height=\outH,
    magnification=\zoomMag,
    connect spies,
    every spy on node/.style ={draw=red, very thick},
    every spy in node/.style ={draw=red, very thick},
  },
]
  \def\D{Figures/diffpir_vs_rd/dmpx_sigma0.1_idx9}

  \node[panel] (phantom) at (0,0)
    {\includegraphics[width=\panelW]{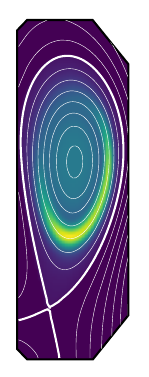}};
  \node[panel] (rdmean)  at ($(phantom.south east)+(\panelSep,0)$)
    {\includegraphics[width=\panelW]{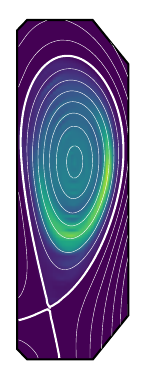}};
  \node[panel] (dpmean)  at ($(rdmean.south east)+(\panelSep,0)$)
    {\includegraphics[width=\panelW]{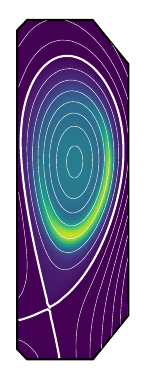}};
  \node[panel, anchor=west] (cbarM) at ($(dpmean.east)+(\cbarSep,0)$)
{\includegraphics[height=0.39\textwidth]{Figures/dataset/phantoms_cbar.pdf}};

  \node[panel] (rdstd)   at ($(cbarM.south east)+(\groupSep,0)$)
    {\includegraphics[width=\panelW]{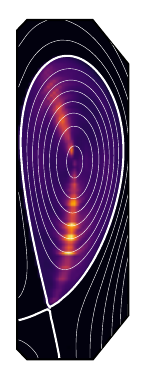}};
  \node[panel] (cbarS1)  at ($(rdstd.south east)+(\cbarSep,0)$)
    {\includegraphics[height=0.39\textwidth]{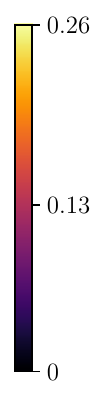}};
  \node[panel, anchor=west] (dpstd)   at ($(cbarS1.east)+(\panelSep,0)$)
    {\includegraphics[width=\panelW]{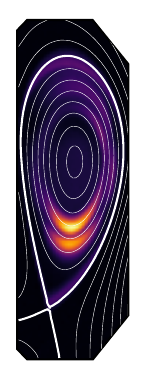}};
  \node[panel, anchor=west] (cbarS2)  at ($(dpstd.east)+(\cbarSep,0)$)
    {\includegraphics[height=0.39\textwidth]{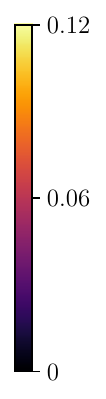}};

\node[anchor=south, font=\small] at ($(phantom.north) + (0, -0.3)$) {$\vx_{_\text{GT}}$};
\node[anchor=south, font=\small] at ($(rdmean.north) + (0, -0.3)$)  {$\vmu_{_\text{RD}}$};
\node[anchor=south, font=\small] at ($(dpmean.north) + (0, -0.3)$)  {$\vmu_{_\text{DiffPIR}}$};
\node[anchor=south, font=\small] at ($(rdstd.north) + (0, -0.3)$)   {$\vsigma_{_\text{RD}}$};
\node[anchor=south, font=\small] at ($(dpstd.north) + (0, -0.3)$)   {$\vsigma_{_\text{DiffPIR}}$};
  \foreach \n in {phantom, rdmean, dpmean}{
    \coordinate (\n-zc) at
      ($(\n.south west)!\zoomCx!(\n.south east)
       !\zoomCy!
       ($(\n.north west)!\zoomCx!(\n.north east)$)$);
  }

  \coordinate (z-phantom) at ($(phantom.south)+(0,-\rowSep-0.25*\outH)$);
  \coordinate (z-rdmean)  at ($(rdmean.south) +(0,-\rowSep-0.25*\outH)$);
  \coordinate (z-dpmean)  at ($(dpmean.south) +(0,-\rowSep-0.25*\outH)$);

  \spy on (phantom-zc) in node at (z-phantom);
  \spy on (rdmean-zc)  in node at (z-rdmean);
  \spy on (dpmean-zc)  in node at (z-dpmean);

\node[anchor=north, font=\small]
  at ($(rdstd.south)!0.5!(dpstd.south) + (0, -\rowSep + 0.7cm)$)
  {%
    \renewcommand{\arraystretch}{1.4}%
    \begin{tabular}{@{}lcc@{}}
    \hline
                              & PSNR           & SSIM           \\
    \hline
    $\vmu_{_\text{RD}}$      & 29.90          & 0.933          \\
    $\vmu_{_\text{DiffPIR}}$ & \textbf{37.08} & \textbf{0.973} \\
    \hline
    \end{tabular}%
  };
\end{tikzpicture}
\captionsetup{justification=justified,singlelinecheck=false}
\caption{DMPX diagnostic: reconstruction of the ground truth $\mathbf{x}_{_{GT}}$, for noise level $\eta=\eta_{low}$. For both RD and DiffPIR, the posterior mean $\vmu$ and standard deviation $\vsigma$ are shown; the value of the performance metrics is also reported.}
\label{fig:diffpir_vs_rd_dmpx}
\end{figure}

We compare DiffPIR with the RD baseline on the SXR and DMPX diagnostics.
For all noise levels, we set DiffPIR hyperparameters to $(\lambda_{pir},\zeta)=(100,0.8)$ for SXR, to $(\lambda_{pir},\zeta)=(300,0.8)$ for DMPX.\\
Figures \ref{fig:diffpir_vs_rd_sxr} and \ref{fig:diffpir_vs_rd_dmpx} show, for one test phantom per diagnostic, the ground truth, the posterior means of the two methods, and their pixel-wise posterior standard deviations, computed as described earlier in the section.
Tables \ref{tab:sxr} and \ref{tab:dmpx} report the average value and standard deviation of the metrics over the full test set.
\\
DiffPIR outperforms RD on both diagnostics and at all noise levels, with average PSNR gains that reach $10\,\mathrm{dB}$ for SXR and $7\,\mathrm{dB}$ for DMPX at the lowest noise level.
The gains shrink as the noise grows, but remain statistically significant in every considered configuration; the single exception is the SSIM at the highest DMPX noise level, where RD is marginally, yet significantly, better.
The magnified insets in Figs.\ref{fig:diffpir_vs_rd_sxr} and \ref{fig:diffpir_vs_rd_dmpx} illustrate the origin of the overall better performance of DiffPIR.
The algorithm reconstructs better than RD complex and fine features of the ground truth, such as the hollow center in Fig.\ref{fig:diffpir_vs_rd_sxr} and the sharp localized structure in Fig.\ref{fig:diffpir_vs_rd_dmpx}, which RD attenuates and blurs, additionally exhibiting some horizontally aligned artifacts.
The shortcomings of RD at reconstructing such details could be observed also in the RD study of \cite{hamm2025}; in particular, they are all clearly visible in Fig.6b of that paper, where the limitations of RD are discussed.
Remarkably, DiffPIR achieves these results without any knowledge of the magnetic equilibrium, on which RD relies.
The flux surfaces overlaid in the figures show that the DiffPIR reconstructions nonetheless remain consistent with the equilibrium: no emissivity is placed outside of the core, and the reconstructed features are aligned with the flux surfaces.
Moreover, as earlier discussed, no DiffPIR hyperparameters were tuned individually for each reconstruction, in contrast with the tuning strategy implemented for RD's hyperparameter $\alpha$.\\ The

\begin{table}[t!]
\centering
\begin{tabular}{llccc}
\toprule
Metric & Method & $\eta_{\text{low}}$ & $\eta_{\text{medium}}$ & $\eta_{\text{high}}$ \\
\midrule
\multirow{2}{*}{PSNR}
 & RD (mean) & $37.1 \pm 3.5$            & $33.9 \pm 3.5$            & $31.1 \pm 3.6$ \\
 & DiffPIR   & $\mathbf{47.1 \pm 3.2}^*$ & $\mathbf{40.0 \pm 3.1}^*$ & $\mathbf{36.0 \pm 3.2}^*$ \\
\midrule
\multirow{2}{*}{SSIM}
 & RD (mean) & $0.97 \pm 0.01$            & $0.95 \pm 0.02$            & $0.93 \pm 0.04$ \\
 & DiffPIR   & $\mathbf{0.99 \pm 0.01}^*$ & $\mathbf{0.98 \pm 0.01}^*$ & $\mathbf{0.97 \pm 0.01}^*$ \\
\bottomrule
\end{tabular}
\captionsetup{justification=justified,singlelinecheck=false}
\caption{SXR reconstruction metrics by method and noise level, reporting average and standard deviation over the test set. Best value in bold; $^*$ denotes that the best-performing method is, statistically, significantly better than the other.}
\label{tab:sxr}
\end{table}

\begin{table}[t!]
\centering
\begin{tabular}{llccc}
\toprule
Metric & Method & $\eta_{\text{low}}$ & $\eta_{\text{medium}}$ & $\eta_{\text{high}}$ \\
\midrule
\multirow{2}{*}{PSNR}
 & RD (mean) & $34.6 \pm 3.6$            & $34.0 \pm 3.2$            & $32.1 \pm 3.2$ \\
 & DiffPIR   & $\mathbf{41.6 \pm 5.2}^*$ & $\mathbf{37.3 \pm 4.7}^*$ & $\mathbf{34.6 \pm 4.2}^*$ \\
\midrule
\multirow{2}{*}{SSIM}
 & RD (mean) & $0.97 \pm 0.01$            & $0.96 \pm 0.02$            & $\mathbf{0.95 \pm 0.02}^*$ \\
 & DiffPIR   & $\mathbf{0.98 \pm 0.01}^*$ & $\mathbf{0.96 \pm 0.02}^*$ & $0.95 \pm 0.02$ \\
\bottomrule
\end{tabular}
\captionsetup{justification=justified,singlelinecheck=false}
\caption{DMPX reconstruction metrics by method and noise level, reporting average and standard deviation over the test set. Best value in bold; $^*$ denotes that the best-performing method is, statistically, significantly better than the other.}
\label{tab:dmpx}
\end{table}

\hspace{-0.65cm}two frameworks differ even more in their uncertainty quantification.
The RD standard deviation spreads over the whole reconstruction region: consistently with the discussion in Section \ref{introduction}, its spatial pattern mainly reflects the structure of the prior, rather than the localization of the actual reconstruction error.
The DiffPIR standard deviation, instead, is smaller in magnitude and concentrates on the steep-gradient features where the posterior mean is found to depart the most from the ground truth.
The map thus carries relevant information: at least for data consistent with the learned prior, large values of $\vsigma_{_\text{DiffPIR}}$ flag the regions where the reconstruction is least reliable.
DiffPIR is thus more informative than RD in terms of uncertainty quantification; this is confirmed by computing the Pearson correlation coefficients $r_{\vsigma, err}$ between the standard deviation $\vsigma$ and the absolute reconstruction error $err=\vert\mathbf{x}_{_{GT}}-\vmu\vert$: for both SXR and DMPX, $r_{\vsigma, err}^{\mathrm{DiffPIR}}$ is $2$-$3$ times larger than $r_{\vsigma, err}^{\mathrm{RD}}$ (depending on the noise level).
It should be noted that the standard deviation, per se, is simply a measure of the dispersion around the mean: even for an ideal model, it is not expected to perfectly correlate with the reconstruction error.
However, the found increase in correlation shows that the enhanced expressiveness of the diffusion prior leads to a model that is more aware of its own limitations, in the sense that it correctly flags as uncertain regions where the actual model reliability is indeed lower.
\\
Moreover, studying the correlation $r_{\vmu, err}$ between the mean and error, we observe that $r_{\vmu, err}^{\mathrm{RD}}\approx r_{\vsigma, err}^{\mathrm{RD}}$, while $r_{\vsigma, err}^{\mathrm{DiffPIR}}$ is significantly larger than $r_{\vmu, err}^{\mathrm{DiffPIR}}$.
This shows that the uncertainty map $\vsigma_{_\text{RD}}$ is not a better error predictor than the reconstruction $\vmu_{_\text{RD}}$ itself, while $\vsigma_{_\text{DiffPIR}}$ carries significantly more structural information about the error than $\vmu_{_\text{DiffPIR}}$.
\\
In terms of model calibration, both RD and DiffPIR are found to be rather conservative: the fraction of pixels that fall within $\vmu\pm\vsigma$ and $\vmu\pm2\vsigma$ is typically well beyond the $68\%$ and $95\%$ marks that would correspond to a well-calibrated Gaussian model.
The considered posteriors are not Gaussian, so exact correspondence to these marks should not be expected, but these values provide a useful reference.
Overall, the obtained empirical coverages at one and two standard deviations confirm that both RD and DiffPIR provide useful operative credible bounds to the emissivity values, as discussed in \cite{hamm2025} for RD.
\\
We further examine the correspondence between the posterior standard deviation and the actual reconstruction error in Section \ref{pilatus}.

\subsection{Phantom study: Pilatus diagnostic}
\label{pilatus}
\begin{figure}[t!]
\centering
\newcommand{\panelW}{0.15\textwidth}
\newcommand{\panelSep}{0.04cm}
\newcommand{\cbarSep}{0.08cm}
\newcommand{\rowSep}{0.5cm}
\begin{tikzpicture}[
  panel/.style={anchor=south west, inner sep=0pt, outer sep=0pt},
]
  \def\D{Figures/pilatus/pilatus_all_sigma0.7_idx52}
  \node[panel] (phantom) at (0,0)
    {\includegraphics[width=\panelW]{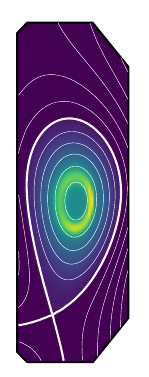}};
  \node[panel] (rdmap) at ($(phantom.south east)+(\panelSep,0)$)
    {\includegraphics[width=\panelW]{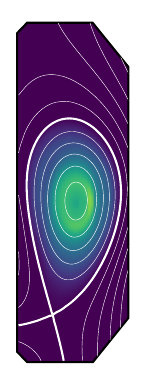}};
  \node[panel] (rdmean) at ($(rdmap.south east)+(\panelSep,0)$)
    {\includegraphics[width=\panelW]{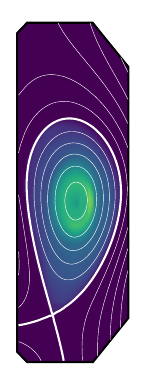}};
  \node[panel] (dpmean) at ($(rdmean.south east)+(\panelSep,0)$)
    {\includegraphics[width=\panelW]{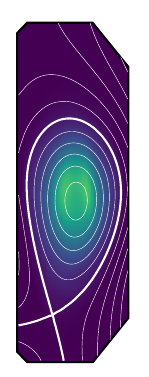}};
  \node[panel] (drdclipmean) at ($(dpmean.south east)+(\panelSep,0)$)
    {\includegraphics[width=\panelW]{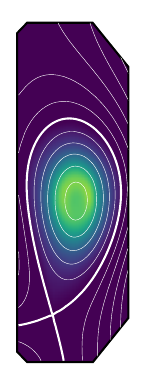}};
  \node[panel] (drdmean) at ($(drdclipmean.south east)+(\panelSep,0)$)
    {\includegraphics[width=\panelW]{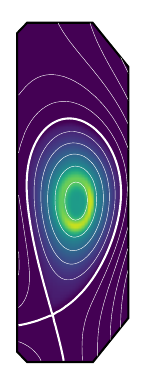}};
  \node[panel, anchor=west] (cbarM) at ($(drdmean.east)+(\cbarSep,0)$)
    {\includegraphics[height=0.39\textwidth]{Figures/dataset/phantoms_cbar.pdf}};

  \node[anchor=south, font=\small] at ($(phantom.north)+(0,-0.3)$)      {$\vx_{_\text{GT}}$};
  \node[anchor=south, font=\small] at ($(rdmap.north)+(0,-0.3)$)        {$\vx_{_\text{RD(MAP)}}$};
  \node[anchor=south, font=\small] at ($(rdmean.north)+(0,-0.3)$)       {$\vmu_{_\text{RD}}$};
  \node[anchor=south, font=\small] at ($(dpmean.north)+(0,-0.3)$)       {$\vmu_{_\text{DiffPIR}}$};
  \node[anchor=south, font=\small] at ($(drdclipmean.north)+(0,-0.3)$)  {$\vmu_{_\text{DiffRD}}^{_\text{clip}}$};
  \node[anchor=south, font=\small] at ($(drdmean.north)+(0,-0.3)$)      {$\vmu_{_\text{DiffRD}}$};

  \node[panel, anchor=north west] (rdstd) at ($(rdmean.south west)+(0,-\rowSep)$)
    {\includegraphics[width=\panelW]{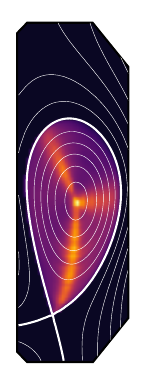}};

  \node[panel, anchor=north west] (dpstd) at ($(dpmean.south west)+(0,-\rowSep)$)
    {\includegraphics[width=\panelW]{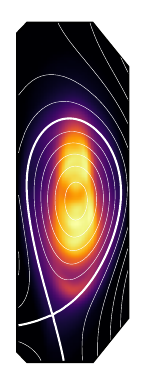}};

  \node[panel, anchor=north west] (drdclipstd) at ($(drdclipmean.south west)+(0,-\rowSep)$)
    {\includegraphics[width=\panelW]{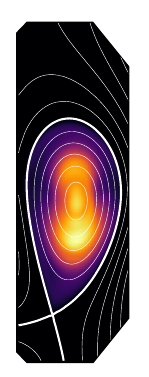}};

  \node[panel, anchor=north west] (drdstd) at ($(drdmean.south west)+(0,-\rowSep)$)
    {\includegraphics[width=\panelW]{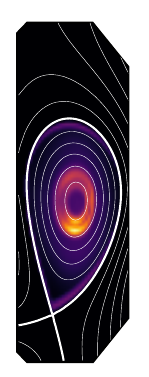}};

  \node[anchor=south, font=\small] at ($(rdstd.north)+(0,-0.3)$)       {$\vsigma_{_\text{RD}}$};
  \node[anchor=south, font=\small] at ($(dpstd.north)+(0,-0.3)$)       {$\vsigma_{_\text{DiffPIR}}$};
  \node[anchor=south, font=\small] at ($(drdclipstd.north)+(0,-0.3)$)  {$\vsigma_{_\text{DiffRD}}^{_{\text{clip}}}$};
  \node[anchor=south, font=\small] at ($(drdstd.north)+(0,-0.3)$)      {$\vsigma_{_\text{DiffRD}}$};

  \newcommand{\hcbarSep}{-0.4cm}
  \node[anchor=north] (hcbarRd) at ($(rdstd.south)+(0,-\hcbarSep)$)
    {\includegraphics[width=\panelW]{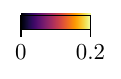}};
  \node[anchor=north] (hcbarDp) at ($(dpstd.south)+(0,-\hcbarSep)$)
    {\includegraphics[width=\panelW]{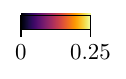}};
  \node[anchor=north] (hcbarDrdClip) at ($(drdclipstd.south)+(0,-\hcbarSep)$)
    {\includegraphics[width=\panelW]{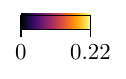}};
  \node[anchor=north] (hcbarDrd) at ($(drdstd.south)+(0,-\hcbarSep)$)
    {\includegraphics[width=\panelW]{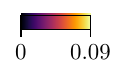}};

  \coordinate (tableX) at ($(phantom.south)!0.5!(rdmap.south)$);
  \node[anchor=north, font=\small]
    at ($(tableX |- rdstd.north)+(0,-0.3cm)$)
    {%
      \renewcommand{\arraystretch}{1.5}%
      \begin{tabular}{@{}l@{\hskip 4pt}c@{\hskip 4pt}c@{}}
      \hline
                                       & PSNR           & SSIM           \\
      \hline
      $\vx_{_\text{RD}}^{_\text{MAP}}$         & 29.50          & 0.910          \\
      $\vmu_{_\text{RD}}$             & 27.16          & 0.903          \\
      $\vmu_{_\text{DiffPIR}}$        & 26.70          & 0.837          \\
      $\vmu_{_\text{DiffRD}}^{_{\text{clip}}}$       & 26.67          & 0.862          \\
      $\vmu_{_\text{DiffRD}}$         & \textbf{33.55} & \textbf{0.943} \\
      \hline
      \end{tabular}%
    };
\end{tikzpicture}
\captionsetup{justification=justified,singlelinecheck=false}
\caption{Pilatus diagnostic: reconstruction of the ground truth $\mathbf{x}_{_{GT}}$, for noise level $\eta=\eta_{high}$. The posterior mean $\vmu$ and standard deviation $\vsigma$ are shown for RD, DiffPIR, DiffRD$_\text{clip}$ and DiffRD. For RD, the MAP $\mathbf{x}_{\mathrm{RD(MAP)}}$ is also shown. The value of the performance metrics is also reported.}
\label{fig:pilatus}
\end{figure}

For the single-camera Pilatus diagnostic, we compare DiffPIR and DiffRD with the RD baseline.
For all noise levels, we set DiffPIR hyperparameters to $(\lambda_{pir},\zeta)=(200,0.8)$.
For DiffRD, we set $(\lambda_{pir},\zeta, \lambda_{rd})=(100,0.9, 10^{-5})$ for $\eta=\eta_{low}$, and $(\lambda_{pir},\zeta, \lambda_{rd})=(10,0.9, 10^{-4})$ for $\eta=\eta_{medium}, \eta_{high}$.
This is the only instance where the diffusion hyperparameters selected by the described grid search exhibit some dependence on the noise level.\\
The single-camera Pilatus diagnostic poses a considerably harder problem than SXR and DMPX.
Its $M=80$ lines of sight all originate from a single camera, which cannot vertically localize the emission along the viewing chords; as discussed in Section \ref{inverse_problems_tomography}, 2D reconstruction is typically not even attempted for this diagnostic.
We nonetheless attempt it, and we compare five reconstructions: the RD posterior mean and MAP, DiffPIR, the proposed DiffRD, which augments the learned prior with the equilibrium-informed RD regularization (Section \ref{posterior_sampling}), and DiffRD$_{\text{clip}}$, an ablation of DiffRD that retains the geometry clipping to the core region but omits the RD term (Algorithm \ref{alg:diffrd}).
The RD MAP estimate $\mathbf{x}_{_{\mathrm{RD(MAP)}}}$ is computed with the gradient ascent in Eq.\eqref{GDULA}; while efficient to compute, it provides no uncertainty quantification.
The MAP is included in the analysis because, while $\mathbf{x}_{_{\mathrm{RD(MAP)}}}$ and $\vmu_{_\text{RD}}$ perform similarly well for SXR and DMPX (as reported for the SXR case in \cite{hamm2025}), the Pilatus diagnostic configuration leads to performance discrepancies between the two.
\\
Figure \ref{fig:pilatus} shows the reconstructions and the posterior standard deviations for one test phantom at the highest noise level; Table \ref{tab:pilatus} reports the metrics over the full test set.
\\
The

\begin{table}[t!]
\centering
\begin{tabular}{llccc}
\toprule
Metric & Method & $\eta_{\text{low}}$ & $\eta_{\text{medium}}$ & $\eta_{\text{high}}$ \\
\midrule
\multirow{5}{*}{PSNR}
 & RD (mean)              & $28.0 \pm 4.2$            & $27.8 \pm 4.0$            & $27.1 \pm 3.9$ \\
 & RD (MAP)               & $29.9 \pm 4.9$            & $29.1 \pm 4.3$            & $28.4 \pm 4.1$ \\
 & DiffPIR                & $22.4 \pm 3.7$            & $22.2 \pm 3.6$            & $22.0 \pm 3.4$ \\
 & DiffRD$_{\text{clip}}$ & $28.9 \pm 5.0$            & $28.2 \pm 4.7$            & $27.5 \pm 4.4$ \\
 & DiffRD                 & $\mathbf{32.9 \pm 6.3}^*$ & $\mathbf{32.1 \pm 6.6}^*$ & $\mathbf{30.9 \pm 6.2}^*$ \\
\midrule
\multirow{5}{*}{SSIM}
 & RD (mean)              & $0.92 \pm 0.05$           & $0.92 \pm 0.05$           & $0.90 \pm 0.05$ \\
 & RD (MAP)               & $0.92 \pm 0.04$           & $0.92 \pm 0.04$           & $0.90 \pm 0.05$ \\
 & DiffPIR                & $0.74 \pm 0.10$           & $0.73 \pm 0.09$           & $0.72 \pm 0.09$ \\
 & DiffRD$_{\text{clip}}$ & $0.91 \pm 0.06$           & $0.89 \pm 0.06$           & $0.88 \pm 0.06$ \\
 & DiffRD                 & $\mathbf{0.95 \pm 0.03}^*$ & $\mathbf{0.95 \pm 0.04}^*$ & $\mathbf{0.94 \pm 0.04}^*$ \\
\bottomrule
\end{tabular}
\captionsetup{justification=justified,singlelinecheck=false}
\caption{Pilatus reconstruction metrics by method and noise level, reporting average and standard deviation over the test set. Best value in bold; $^*$ denotes that the best-performing method is, statistically, significantly better than the other.}
\label{tab:pilatus}
\end{table}

\begin{figure}[t!]
\centering
\newcommand{\panelW}{0.15\textwidth}
\newcommand{\panelSep}{-0.3cm}
\newcommand{\groupSep}{0.6cm}
\newcommand{\rowSep}{0.1cm}
\begin{tikzpicture}[
  panel/.style={anchor=south west, inner sep=0pt, outer sep=0pt},
]
  \def\Da{Figures/std_vs_err/pilatus_all_sigma0.35_idx666}
  \def\Db{Figures/std_vs_err/pilatus_all_sigma0.35_idx608}
  \def\Dc{Figures/std_vs_err/pilatus_all_sigma0.35_idx969}

  \node[panel] (rdstdA) at (0,0)
    {\includegraphics[width=\panelW]{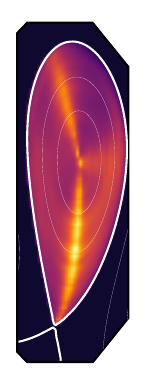}};
  \node[panel] (rderrA) at ($(rdstdA.south east)+(\panelSep,0)$)
    {\includegraphics[width=\panelW]{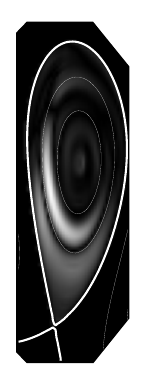}};
  \node[panel] (rdstdB) at ($(rderrA.south east)+(\groupSep,0)$)
    {\includegraphics[width=\panelW]{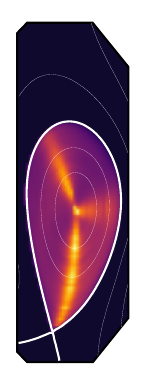}};
  \node[panel] (rderrB) at ($(rdstdB.south east)+(\panelSep,0)$)
    {\includegraphics[width=\panelW]{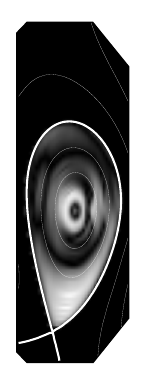}};
  \node[panel] (rdstdC) at ($(rderrB.south east)+(\groupSep,0)$)
    {\includegraphics[width=\panelW]{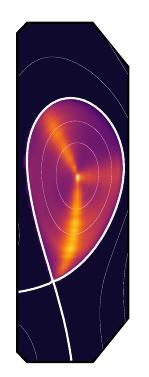}};
  \node[panel] (rderrC) at ($(rdstdC.south east)+(\panelSep,0)$)
    {\includegraphics[width=\panelW]{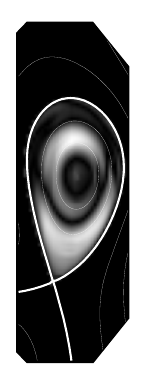}};

  \node[anchor=south, font=\small] at ($(rdstdA.north)+(0,-0.3)$)  {$\vsigma$};
  \node[anchor=south, font=\small] at ($(rderrA.north)+(0,-0.3)$)  {$err$};
  \node[anchor=south, font=\small] at ($(rdstdB.north)+(0,-0.3)$)  {$\vsigma$};
  \node[anchor=south, font=\small] at ($(rderrB.north)+(0,-0.3)$)  {$err$};
  \node[anchor=south, font=\small] at ($(rdstdC.north)+(0,-0.3)$)  {$\vsigma$};
  \node[anchor=south, font=\small] at ($(rderrC.north)+(0,-0.3)$)  {$err$};

  \node[panel, anchor=north west] (drdstdA) at ($(rdstdA.south west)+(0,-\rowSep)$)
    {\includegraphics[width=\panelW]{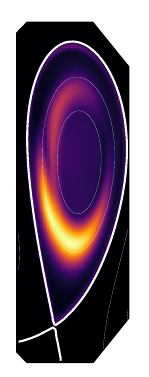}};
  \node[panel, anchor=north west] (drderrA) at ($(rderrA.south west)+(0,-\rowSep)$)
    {\includegraphics[width=\panelW]{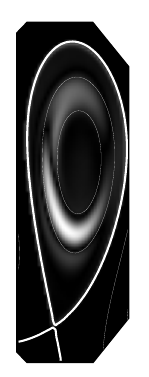}};
  \node[panel, anchor=north west] (drdstdB) at ($(rdstdB.south west)+(0,-\rowSep)$)
    {\includegraphics[width=\panelW]{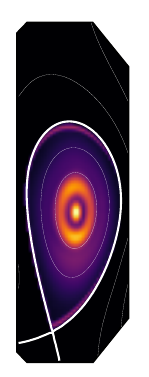}};
  \node[panel, anchor=north west] (drderrB) at ($(rderrB.south west)+(0,-\rowSep)$)
    {\includegraphics[width=\panelW]{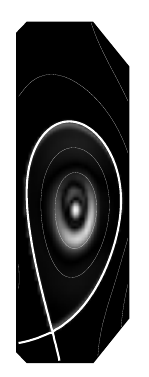}};
  \node[panel, anchor=north west] (drdstdC) at ($(rdstdC.south west)+(0,-\rowSep)$)
    {\includegraphics[width=\panelW]{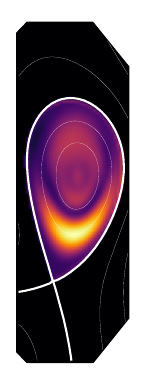}};
  \node[panel, anchor=north west] (drderrC) at ($(rderrC.south west)+(0,-\rowSep)$)
    {\includegraphics[width=\panelW]{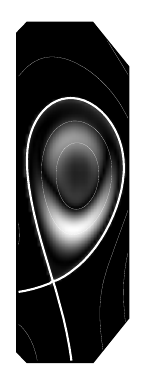}};

  \node[anchor=east, font=\small, rotate=90] at ($(rdstdA.west)!0.5!(rdstdA.north west)-(0.4cm,0)$)
    {RD};
  \node[anchor=east, font=\small, rotate=90] at ($(drdstdA.west)!0.5!(drdstdA.north west)-(0.4cm,0)$)
    {DiffRD};

\end{tikzpicture}
\captionsetup{justification=justified,singlelinecheck=false}
\caption{Posterior standard deviations vs.\ absolute errors for three reconstructions computed for the Pilatus diagnostic. Top: RD; bottom: DiffRD. Columns: standard deviation $\vsigma$ and reconstruction error $err=\vert\mathbf{x}_{_{GT}}-\vmu\vert$ of the posterior mean $\vmu$.}
\label{fig:std_vs_err}
\end{figure}
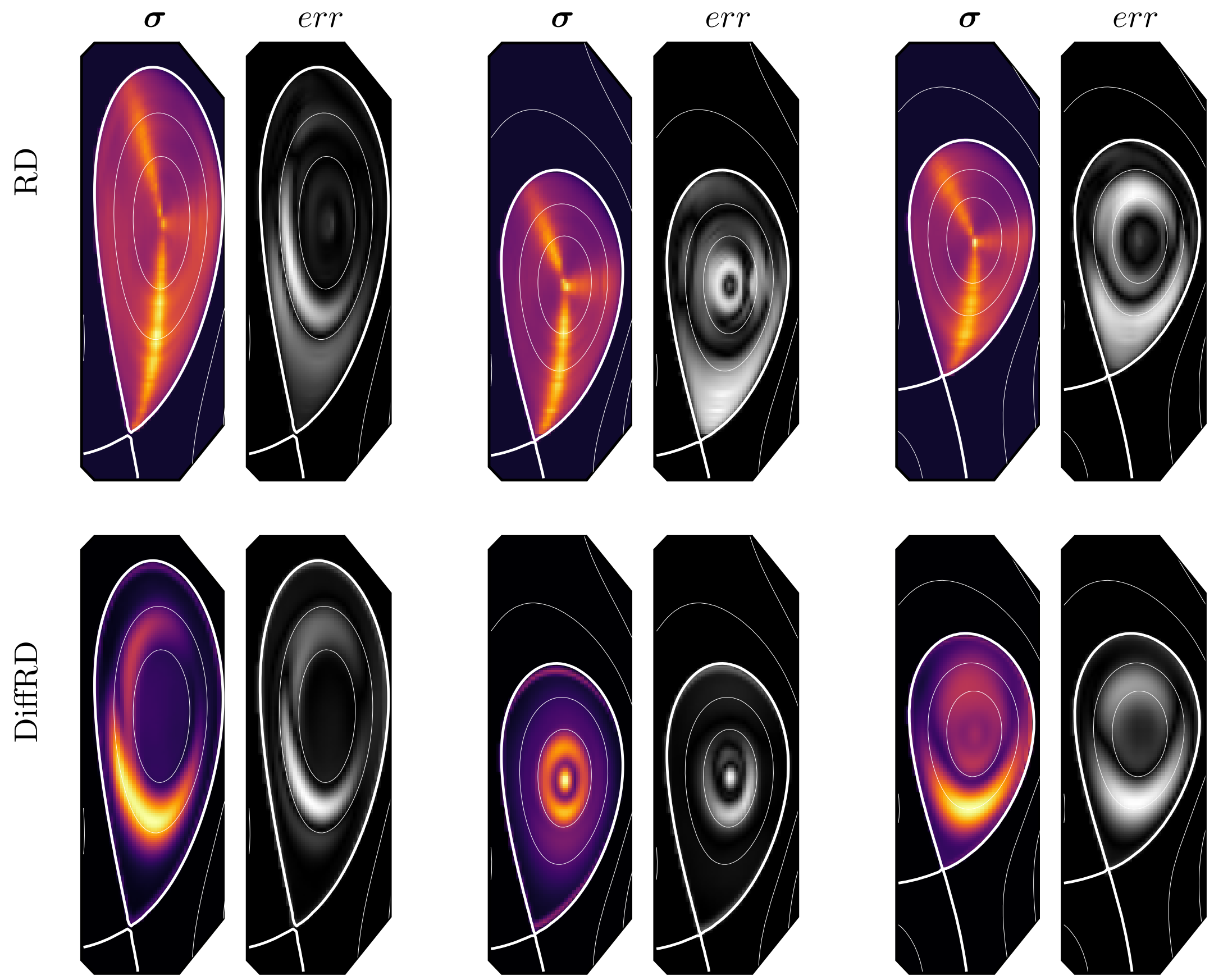

\hspace{-0.35cm}results in Table \ref{tab:pilatus} confirm the difficulty of the problem and the value of the equilibrium information.
Plain DiffPIR fails: its PSNR falls $5$ to $6\,\mathrm{dB}$ below that of the RD posterior mean, showing that the learned prior alone cannot compensate for the missing views.
The geometry clipping recovers most of the gap, with DiffRD$_{\text{clip}}$ performing on par with the RD posterior mean and slightly worse than the RD MAP.
The full DiffRD outperforms every competitor at all noise levels, with statistically significant margins: it gains about $3\,\mathrm{dB}$ in PSNR over RD (MAP), the strongest baseline, and raises the SSIM from $\approx0.92$ to $\approx0.95$.
For the RD baseline itself, we note that the MAP slightly outperforms the posterior mean on this diagnostic.
\\
The posterior means and standard deviations in Fig.\ref{fig:pilatus} explain the ranking of the diffusion-based methods.
The DiffPIR posterior struggles to assess the vertical position of the plasma, unable to localize the emissivity along the chords.
The clipping in DiffRD$_{\text{clip}}$ improves the situation, but the predicted uncertainty is still spread over a large section of the core; emissivity features are typically smoothed out.
The equilibrium information injected by DiffRD contracts the posterior, whose standard deviation scale is considerably smaller than that of the other methods; complex emissivity features are reconstructed with great accuracy.
\\
Remarkably, DiffRD outperforms both DiffPIR and RD: the method inherits the best features of the two methods it is derived from, combining them into a more accurate reconstruction model.
In general, DiffRD reconstructs extremely well emissivity patterns that are Gaussian-like, peaked, hollow, or exhibiting in/out poloidal asymmetries, while top/bottom ones are not always well recovered.
However, even for the phantoms whose reconstruction is poorer, the method does not perform worse than its components: these simply correspond to profiles whose features are genuinely hard to recover given the diagnostic layout.
Overall, the results demonstrate that meaningful 2D reconstructions can be obtained also for the single-camera Pilatus, without having to constrain the emissivity to be constant on flux surfaces.
\\
Figure \ref{fig:std_vs_err} presents the analysis anticipated in Section \ref{sxr_vs_dmpx}: for three test phantoms, it compares the standard deviation $\vsigma$ with the absolute reconstruction error $err$ of the posterior mean, for both RD and DiffRD.
For RD, the two maps show little correspondence, and the standard deviation displays a similar, prior-driven pattern for all phantoms.
For DiffRD, the standard deviation reproduces rather closely the spatial structure of the error, phantom by phantom.
The map thus constitutes a meaningful uncertainty quantification: without access to the ground truth, it indicates where the reconstruction is least reliable.
\\
Similarly to what we discussed for SXR and DMPX, $r_{\vsigma, err}^{\mathrm{DiffRD}}$ is approximately $2$ times larger than $r_{\vsigma, err}^{\mathrm{RD}}$ and $r_{\vmu, err}^{\mathrm{DiffRD}}$ for all noise levels, while $r_{\vsigma, err}^{\mathrm{RD}}\approx r_{\vmu, err}^{\mathrm{RD}}$.
Interestingly $r_{\vsigma, err}^{\mathrm{DiffPIR}}$ is also rather large, only $\approx10$-$15\%$ smaller than $r_{\vsigma, err}^{\mathrm{DiffRD}}$; this shows that DiffPIR remains mostly aware of its limitations even when it performs rather poorly, as in the case of the considered Pilatus diagnostic.
Finally, in terms of empirical coverage at one and two standard deviations, we find that RD almost perfectly matches the earlier discussed Gaussian marks.
DiffRD, instead, is overconfident in its predictions: while $\vsigma_{_\text{DiffRD}}$ well predicts the spatial pattern of the error, its scale underestimates the reconstruction error; its credible intervals thus need to be widened to obtain meaningful operative bounds to the emissivity values.

\subsection{Experimental data: SXR diagnostic}
\label{experimental_data}
\begin{figure}[t!]
\centering
\newcommand{\panelW}{0.1425\textwidth}
\newcommand{\panelSep}{-0.2cm}
\newcommand{\rowSep}{0.3cm}
\begin{tikzpicture}[
  panel/.style={anchor=south west, inner sep=0pt, outer sep=0pt},
]
  \def\D{Figures/real_data}
  \hspace{-0.1cm}
  \node[panel] (rdA) at (0,0)
    {\includegraphics[width=\panelW]{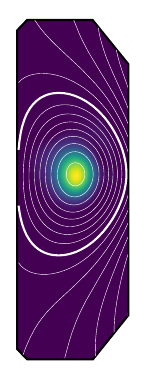}};
  \node[panel] (rdB) at ($(rdA.south east)+(\panelSep,0)$)
    {\includegraphics[width=\panelW]{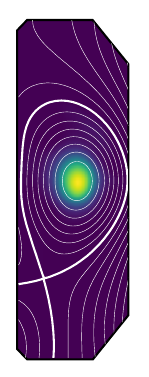}};
  \node[panel] (rdC) at ($(rdB.south east)+(\panelSep,0)$)
    {\includegraphics[width=\panelW]{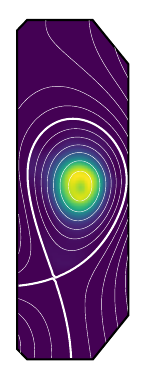}};
  \node[panel] (rdD) at ($(rdC.south east)+(\panelSep,0)$)
    {\includegraphics[width=\panelW]{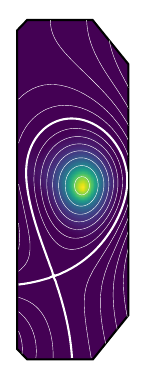}};
  \node[panel] (rdE) at ($(rdD.south east)+(\panelSep,0)$)
    {\includegraphics[width=\panelW]{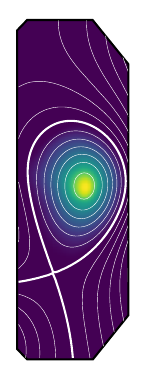}};
  \node[panel] (rdF) at ($(rdE.south east)+(\panelSep,0)$)
    {\includegraphics[width=\panelW]{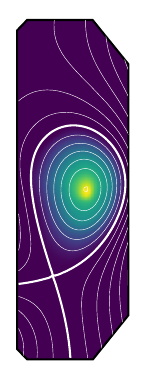}};
  \node[panel] (rdG) at ($(rdF.south east)+(\panelSep,0)$)
    {\includegraphics[width=\panelW]{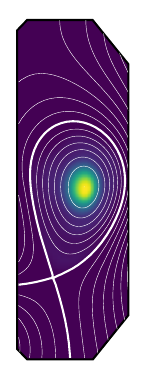}};
     \node[panel, anchor=west] (cbarM) at ($(rdG.east)+(-0.2,0)$)
    {\includegraphics[height=0.39\textwidth]{Figures/dataset/phantoms_cbar.pdf}};

  \node[panel, anchor=north west] (dpA) at ($(rdA.south west)+(0,-\rowSep)$)
    {\includegraphics[width=\panelW]{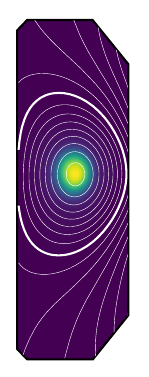}};
  \node[panel, anchor=north west] (dpB) at ($(rdB.south west)+(0,-\rowSep)$)
    {\includegraphics[width=\panelW]{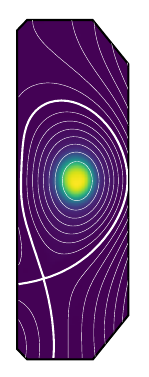}};
  \node[panel, anchor=north west] (dpC) at ($(rdC.south west)+(0,-\rowSep)$)
    {\includegraphics[width=\panelW]{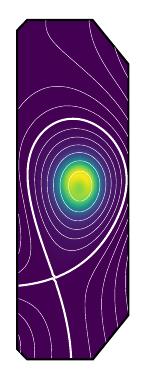}};
  \node[panel, anchor=north west] (dpD) at ($(rdD.south west)+(0,-\rowSep)$)
    {\includegraphics[width=\panelW]{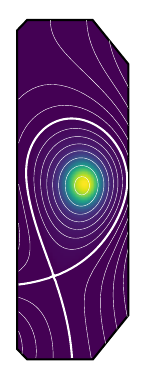}};
  \node[panel, anchor=north west] (dpE) at ($(rdE.south west)+(0,-\rowSep)$)
    {\includegraphics[width=\panelW]{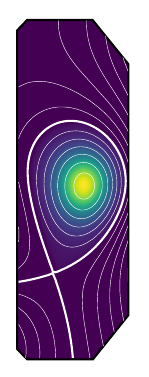}};
  \node[panel, anchor=north west] (dpF) at ($(rdF.south west)+(0,-\rowSep)$)
    {\includegraphics[width=\panelW]{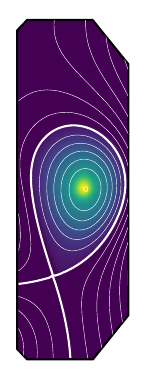}};
  \node[panel, anchor=north west] (dpG) at ($(rdG.south west)+(0,-\rowSep)$)
    {\includegraphics[width=\panelW]{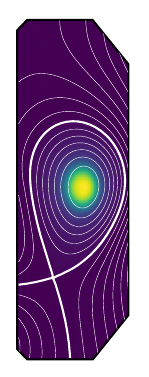}};
     \node[panel, anchor=west] (cbarM2) at ($(dpG.east)+(-0.2,0)$)
    {\includegraphics[height=0.39\textwidth]{Figures/dataset/phantoms_cbar.pdf}};

  \node[anchor=south, font=\small] at ($(rdA.north)+(0,-0.3)$) {$t=0.24\,$s};
  \node[anchor=south, font=\small] at ($(rdB.north)+(0,-0.3)$) {$t=0.42\,$s};
  \node[anchor=south, font=\small] at ($(rdC.north)+(0,-0.3)$) {$t=0.66\,$s};
  \node[anchor=south, font=\small] at ($(rdD.north)+(0,-0.3)$) {$t=0.72\,$s};
  \node[anchor=south, font=\small] at ($(rdE.north)+(0,-0.3)$) {$t=0.78\,$s};
  \node[anchor=south, font=\small] at ($(rdF.north)+(0,-0.3)$) {$t=0.90\,$s};
  \node[anchor=south, font=\small] at ($(rdG.north)+(0,-0.3)$) {$t=1.02\,$s};

  \node[anchor=east, font=\small, rotate = 90] at ($(rdA.west)+(-0.15,0.8)$) {$\vx_{_\text{RD(MAP)}}$};
  \node[anchor=east, font=\small, rotate = 90] at ($(dpA.west)+(-0.15,0.8)$) {$\vmu_{_\text{DiffPIR}}$};
\end{tikzpicture}
\captionsetup{justification=justified,singlelinecheck=false}
\caption{SXR reconstructions of TCV discharge 85270 at different times. Top row: RD MAP
estimates $\mathbf{x}_{_{\mathrm{RD(MAP)}}}$; bottom row: DiffPIR posterior means $\vmu_{_{\mathrm{DiffPIR}}}$.}
\label{fig:real_data_85270}
\end{figure}

\begin{figure}[t!]
\centering
\newcommand{\panelW}{0.1425\textwidth}
\newcommand{\panelSep}{-0.02cm}
\newcommand{\panelSepa}{0.5cm}
\begin{tikzpicture}[
  panel/.style={anchor=south west, inner sep=0pt, outer sep=0pt},
]
  \def\D{Figures/real_data_posterior_samples}
  \hspace{-0.1cm}
    \node[panel] (smpstd) at (0,0)
    {\includegraphics[width=\panelW]{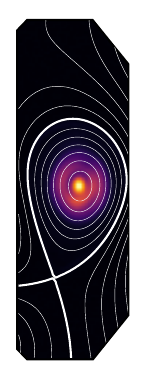}};
    \node[panel] (smpA) at ($(smpstd.south east)+(\panelSepa,0)$)
    {\includegraphics[width=\panelW]{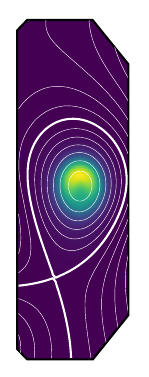}};
  \node[panel] (smpB) at ($(smpA.south east)+(\panelSep,0)$)
    {\includegraphics[width=\panelW]{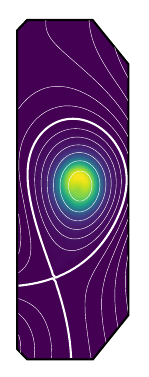}};
  \node[panel] (smpC) at ($(smpB.south east)+(\panelSep,0)$)
    {\includegraphics[width=\panelW]{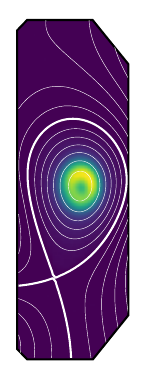}};
  \node[panel] (smpD) at ($(smpC.south east)+(\panelSep,0)$)
    {\includegraphics[width=\panelW]{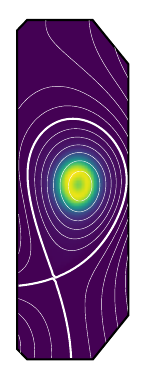}};
  \node[panel] (smpF) at ($(smpD.south east)+(\panelSep,0)$)
    {\includegraphics[width=\panelW]{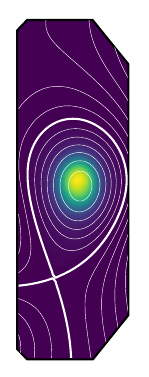}};
  \node[panel] (smpG) at ($(smpF.south east)+(\panelSep,0)$)
    {\includegraphics[width=\panelW]{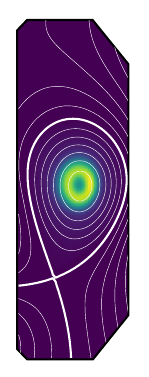}};

  \node[anchor=south, font=\small] at ($(smpstd.north)+(0,-0.3)$) {$\vsigma_{_\text{DiffPIR}}$};
\node[anchor=south, font=\small] at ($(smpA.north)+(0,-0.3)$) {$\mathbf{x}^{(1)}_{_\text{DiffPIR}}$};
\node[anchor=south, font=\small] at ($(smpB.north)+(0,-0.3)$) {$\mathbf{x}^{(2)}_{_\text{DiffPIR}}$};
\node[anchor=south, font=\small] at ($(smpC.north)+(0,-0.3)$) {$\mathbf{x}^{(3)}_{_\text{DiffPIR}}$};
\node[anchor=south, font=\small] at ($(smpD.north)+(0,-0.3)$) {$\mathbf{x}^{(4)}_{_\text{DiffPIR}}$};
\node[anchor=south, font=\small] at ($(smpF.north)+(0,-0.3)$) {$\mathbf{x}^{(5)}_{_\text{DiffPIR}}$};
\node[anchor=south, font=\small] at ($(smpG.north)+(0,-0.3)$) {$\mathbf{x}^{(6)}_{_\text{DiffPIR}}$};

\end{tikzpicture}
\captionsetup{justification=justified,singlelinecheck=false}
\caption{DiffPIR reconstruction of SXR data, for TCV discharge 85270 at $t=0.66\,$s: posterior standard deviation $\vsigma_{_\text{DiffPIR}}$ and six posterior samples $\{\mathbf{x}_{_\text{DiffPIR}}^{(i)}\}_{i=1}^6$.}
\label{fig:real_data_posterior_samples}
\end{figure}
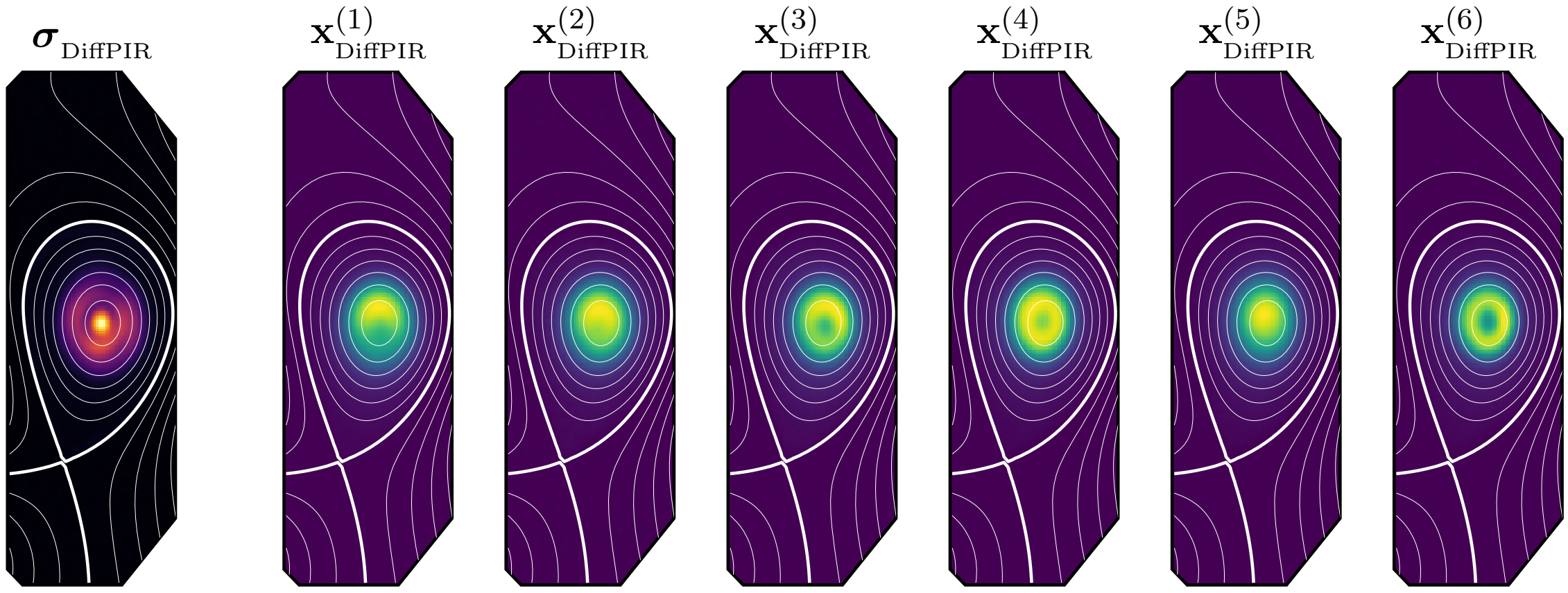

In this section, we apply the equilibrium-free DiffPIR to experimental TCV data from the SXR diagnostic, comparing the results with the RD baseline.
For RD, we use the same hyperparameter strategy used in the previous sections, estimating $\alpha$ individually for each reconstruction.
For DiffPIR, we use the same hyperparameters identified in the SXR phantom study in Section \ref{sxr_vs_dmpx}.
For both, we assume a noise level $\eta=\eta_{medium}$.
\\
Compared to the phantoms considered so far, whose dynamic range varied little across different phantoms, experimental data exhibit amplitudes that depend on the emissivity level, spanning multiple orders of magnitude.
As a consequence, the tomographic data need some pre-processing: the value of $\lambda$ in RD can be treated as a fixed scale parameter only upon data normalization \cite{hamm2026a}, and the performance of the learned diffusion prior considerably degrades if applied to data whose dynamic range is significantly different from that of the training samples.
To account for this, the experimental data $\mathbf{y}_{exp}\in\mathbb{R}^M$ are normalized by a factor $c^*$ defined as
\begin{equation}\label{normalization_equation}
    c^* = \argmin_{c\in\mathbb{R}} \Big\Vert \mathbf{y}^* - \frac{\mathbf{y}_{exp}}{c}\Big\Vert_2^2\;,\quad \mathrm{with}\quad
    \mathbf{y}^* = \argmin_{\mathbf{y}\in\mathcal{Y}_{train}} \Big\Vert \frac{\mathbf{y}}{\Vert\mathbf{y}\Vert_2^2} - \frac{\mathbf{y}_{exp}}{\Vert\mathbf{y}_{exp}\Vert_2^2}\Big\Vert_2^2\;;
\end{equation}
the data are thus normalized to $\widehat{\mathbf{y}}_{exp}=\mathbf{y}_{exp}/c^*$, where the factor is determined as the one that minimizes the residual between $\widehat{\mathbf{y}}_{exp}$ and $\mathbf{y}^*$, with the latter identified as the closest pattern to $\mathbf{y}_{exp}$ among the tomographic data $\mathcal{Y}_{train}=\{\mathbf{y}_{train}^{i}\}_{i=1}^{10^4}$ computed from $10^4$ of the samples used to train the diffusion model.
Essentially, the normalization factor is determined to bring the data close to the scale of the training samples, to avoid being excessively far from it.
Phantom studies conducted on RD and DiffPIR confirmed that the described procedure achieves satisfactory results; the performance of DiffPIR degrades more than that of RD, but the performance margins in favor of DiffPIR remain significant.
\\
In Fig.\ref{fig:real_data_85270}, we show the SXR reconstructions computed for TCV discharge 85270 at different times.
Qualitatively, the RD and DiffPIR reconstructions appear similar and consistent between each other in terms of overall pattern and peakedness, showing that the learned diffusion model can perform well also on experimental data.
As remarked also in the phantom studies, the profiles computed with the equilibrium-free DiffPIR are consistent with the underlying magnetic equilibrium.
In this sense, DiffPIR yields an independent verification of the equilibrium reconstruction: RD is biased by its anisotropic equilibrium-informed prior, which would force consistency with the prescribed equilibrium regardless of the equilibrium's correctness.
\\
Both methods suggest that the soft X-ray emissivity at $t=0.66\mathrm{s}$, unlike the other profiles in Fig.\ref{fig:real_data_85270}, does not peak on the magnetic axis: RD reconstructs a hollow ring-like feature, DiffPIR a top/bottom asymmetry.
From RD alone, it is unclear how to further assess whether the reconstructed feature is effectively present or due to an artifact.
In principle, it could be partially due to an ill-estimated value of the anisotropic parameter $\alpha$: phantom analyses show that an excessively aggressive anisotropic smoothing can sometimes lead to such behavior.
DiffPIR, instead, allows the direct inspection of multiple posterior samples, which represent different plausible solutions to the observed tomographic data.\\
In Fig.\ref{fig:real_data_posterior_samples}, we show the DiffPIR standard deviation and six posterior samples for the SXR reconstruction at $t=0.66\mathrm{s}$.

\begin{figure}[t!]
\centering
\newcommand{\panelW}{0.15\textwidth}
\newcommand{\panelSep}{-0.2cm}
\newcommand{\groupSep}{0.65cm}
\newcommand{\rowSep}{-0.cm}
\begin{tikzpicture}[
  panel/.style={anchor=south west, inner sep=0pt, outer sep=0pt},
]
  \def\D{Figures/real_data}
  \node[panel] (rdBg) at (0,0)
    {\includegraphics[width=\panelW]{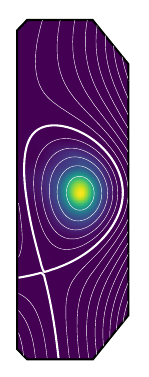}};
  \node[panel] (rdA) at ($(rdBg.south east)+(\groupSep,0)$)
    {\includegraphics[width=\panelW]{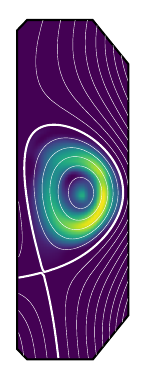}};
  \node[panel] (rdB) at ($(rdA.south east)+(\panelSep,0)$)
    {\includegraphics[width=\panelW]{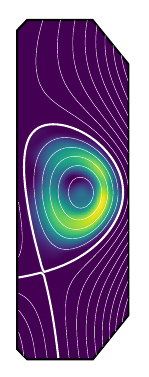}};
  \node[panel] (rdC) at ($(rdB.south east)+(\panelSep,0)$)
    {\includegraphics[width=\panelW]{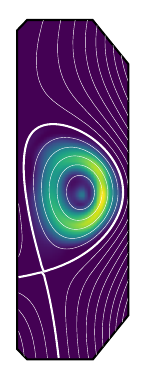}};
  \node[panel] (rdD) at ($(rdC.south east)+(\panelSep,0)$)
    {\includegraphics[width=\panelW]{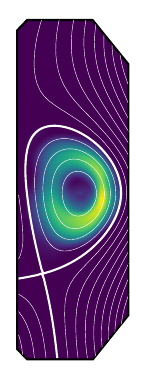}};
  \node[panel] (rdE) at ($(rdD.south east)+(\panelSep,0)$)
    {\includegraphics[width=\panelW]{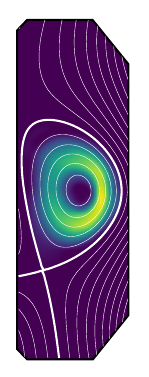}};
     \node[panel, anchor=west] (cbarM) at ($(rdE.east)+(0.05,0)$)
    {\includegraphics[height=0.39\textwidth]{Figures/dataset/phantoms_cbar.pdf}};

  \node[panel, anchor=north west] (dpBg) at ($(rdBg.south west)+(0,-\rowSep)$)
    {\includegraphics[width=\panelW]{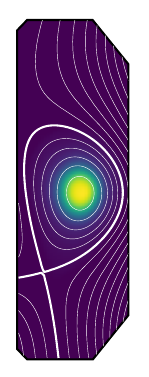}};
  \node[panel, anchor=north west] (dpA) at ($(rdA.south west)+(0,-\rowSep)$)
    {\includegraphics[width=\panelW]{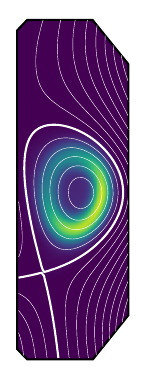}};
  \node[panel, anchor=north west] (dpB) at ($(rdB.south west)+(0,-\rowSep)$)
    {\includegraphics[width=\panelW]{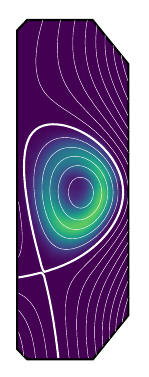}};
  \node[panel, anchor=north west] (dpC) at ($(rdC.south west)+(0,-\rowSep)$)
    {\includegraphics[width=\panelW]{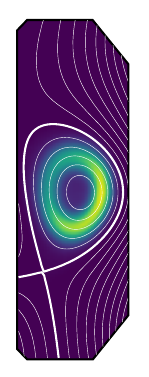}};
  \node[panel, anchor=north west] (dpD) at ($(rdD.south west)+(0,-\rowSep)$)
    {\includegraphics[width=\panelW]{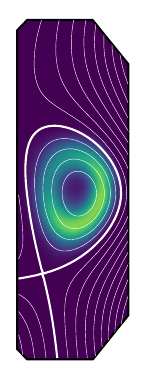}};
  \node[panel, anchor=north west] (dpE) at ($(rdE.south west)+(0,-\rowSep)$)
    {\includegraphics[width=\panelW]{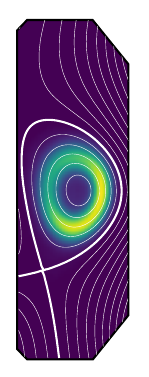}};
     \node[panel, anchor=west] (cbarM) at ($(dpE.east)+(0.05,0)$)
    {\includegraphics[height=0.39\textwidth]{Figures/dataset/phantoms_cbar.pdf}};

  \node[anchor=south, font=\small, align=center] at ($(rdBg.north)+(0,-0.3)$)
    {background\\
    \vspace{-0.35cm}\\
    $t\!=\!1.000\,$s};
\node[anchor=south, font=\small, align=center] at ($(rdC.north)+(0,0.3)$)
    {fluctuations};
  \node[anchor=south, font=\small] at ($(rdA.north)+(0,-0.3)$) {$t\!=\!1.037\,$s};
  \node[anchor=south, font=\small] at ($(rdB.north)+(0,-0.3)$) {$t\!=\!1.038\,$s};
  \node[anchor=south, font=\small] at ($(rdC.north)+(0,-0.3)$) {$t\!=\!1.039\,$s};
  \node[anchor=south, font=\small] at ($(rdD.north)+(0,-0.3)$) {$t\!=\!1.040\,$s};
  \node[anchor=south, font=\small] at ($(rdE.north)+(0,-0.3)$) {$t\!=\!1.041\,$s};

  \node[anchor=east, font=\small, rotate=90] at ($(rdBg.west)+(-0.15,0.8)$) {$\vx_{_\text{RD(MAP)}}$};
  \node[anchor=east, font=\small, rotate=90] at ($(dpBg.west)+(-0.15,0.8)$) {$\vmu_{_\text{DiffPIR}}$};
\end{tikzpicture}
\captionsetup{justification=justified,singlelinecheck=false}
\caption{TCV discharge 88767: reconstructed in/out soft X-ray asymmetry following N2 injection. First column: background profile at $t=1\,$s; remaining columns:
background-subtracted fluctuation reconstructions at consecutive $1\,$ms
frames. Top row: RD MAP estimates $\mathbf{x}_{_\text{RD(MAP)}}\,$; bottom row: DiffPIR posterior means $\vmu_{_\text{DiffPIR}}$.}
\label{fig:real_data_88767_fluct}
\end{figure}

\hspace{-0.5cm}The DiffPIR posterior samples and standard deviation, which was shown in the previous sections to be a good indicator of model reliability, provide relevant complementary information beyond what can be conveyed by a single point estimate.
In this example, $\vsigma_{_\text{DiffPIR}}$ indicates that the uncertainty in the reconstruction is highest in correspondence of the magnetic axis.
The posterior samples exhibit features that range from hollow profiles to top/bottom asymmetries and even, in some cases, Gaussian-like patterns with a peak close to the magnetic axis: these are all genuinely possible solutions given the observed data.
The extent to which a given feature is likely to be effectively present can be assessed from the proportion of samples exhibiting that pattern.
In the considered case, while the actual emissivity pattern cannot be conclusively determined, the vast proportion of samples that do not peak on the magnetic axis suggests that the true underlying profile, highly likely, does not either.
For RD, posterior sampling by ULA would provide operative credible bounds, but carry little information in terms of error localization, as discussed in the previous sections.
Moreover, while the empirical moments computed from the ULA samples well approximate those of the RD posterior (mean, variance, etc.), inspection of the samples themselves shows that these are considerably noisy, and cannot be interpreted as clean profiles that are different possible solutions.
Diffusion methods such as DiffPIR provide new tools for the statistical analysis of the underdetermined plasma tomography problem.\\
To conclude this section, we compare RD with DiffPIR on SXR data from TCV discharge 88767; the obtained results are shown in Fig.\ref{fig:real_data_88767_fluct}.

In this experiment, N2 was injected from a gas valve located at the inner mid-plane (high-field side), starting from $t=1\,\mathrm{s}$.
We compute the background emissivity level by considering the time-averaged tomographic data over a short time window of $5\,\mathrm{ms}$ around injection time.
Then, we subtract the time-averaged SXR measurements from the signal for $t\in[1\,\mathrm{s},1.05\,\mathrm{s}]$, and reconstruct the resulting background-subtracted emissivity profiles, which thus represent fluctuations around the background.
In the figure, we report the reconstructed background, and the fluctuations computed at $1\,\mathrm{kHz}$ rate for $t\in[1.037\,\mathrm{s},1.041\,\mathrm{s}]$.
The reconstructed fluctuations clearly indicate the presence of an in/out asymmetry in the soft X-ray radiation, observed during a small transitory phase that we attribute to the time at which a significant amount of gas enters into the core.
For $t=1.037$-$1.039\,\mathrm{s}$, the RD reconstructions show a dim radiation feature localized around the magnetic axis, which is absent in $\vmu_{_\text{DiffPIR}}$; this appears consistent with the observations from Fig.\ref{fig:diffpir_vs_rd_sxr}, whose magnified insets showed that DiffPIR successfully removed a similar RD artifact.

\subsection{Limitations: out-of-distribution data}
\label{limitations}
To complete our discussion, we address here some of the possible limitations of the proposed diffusion framework.
Indeed, while the physics-based likelihood guidance embedded in the sampling algorithm actively promotes solutions that are compatible with the observed tomographic data, performance is expected to degrade for profiles that are significantly different from the training samples.
The diffusion prior tends to bias solutions towards profiles that are ``known’’ to it.
This downside is not diffusion-specific, but rather common to all data-driven approaches; it is the price to pay for the increased model expressiveness.
In this section, we consider two numerical experiments designed to illustrate and investigate this potential issue; the results are shown in Fig.\ref{fig:limitations}.
\\
In the top row of Fig.\ref{fig:limitations}, we compare RD with DiffPIR on the SXR reconstruction of a synthetic emissivity profile of a doublet; this plasma configuration, which features two magnetic axes and plasma lobes, is expected to yield soft X-ray profiles similar to the shown phantom $\mathbf{x}_{_\text{GT}}$.
Remarkably, while the diffusion prior was trained on singlet plasmas only, DiffPIR manages to correctly recover two distinct radiation lobes.
Performance metrics reveal that the DiffPIR posterior mean $\vmu_{_\text{DiffPIR}}$ is slightly more accurate than $\vmu_{_\text{RD}}$; it is, however, less accurate than $\mathbf{x}_{_\text{RD(MAP)}}$ (metrics not shown).
\\
\newcommand{\panelW}{0.145\textwidth}
\newcommand{\panelSep}{-0.15cm}
\newcommand{\groupSep}{0.9cm}
\newcommand{\cbarW}{0.145\textwidth}
\newcommand{\cbarSep}{0.1cm}
\newcommand{\labSep}{0.2}
\newcommand{\rowSep}{5.8cm}
\newcommand{\rowLabSep}{0.15cm}
\tikzset{
  panel/.style={anchor=south west, inner sep=0pt, outer sep=0pt},
  cbar/.style={anchor=north, inner sep=0pt, outer sep=0pt},
  lab/.style={anchor=south, font=\small, inner sep=0pt, outer sep=0pt},
  rowlab/.style={rotate=90, anchor=south, font=\small, inner sep=0pt, outer sep=0pt},
}
\begin{figure}[t!]
\centering
\begin{tikzpicture}
  \def\D{Figures/limitations}
  \node[panel] (Dgt) at (0,0)
    {\includegraphics[width=\panelW]{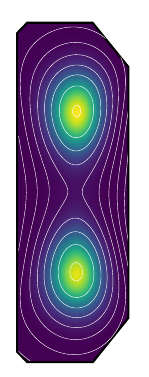}};
  \node[panel] (Drdmap) at ($(Dgt.south east)+(\groupSep,0)$)
    {\includegraphics[width=\panelW]{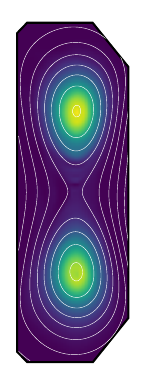}};
  \node[panel] (Drdmean) at ($(Drdmap.south east)+(\panelSep,0)$)
    {\includegraphics[width=\panelW]{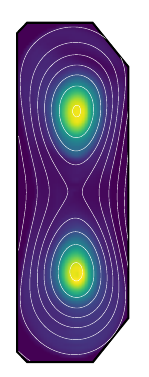}};
  \node[panel] (Drdstd) at ($(Drdmean.south east)+(\panelSep,0)$)
    {\includegraphics[width=\panelW]{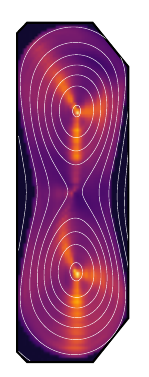}};
  \node[panel] (Ddpmean) at ($(Drdstd.south east)+(\groupSep,0)$)
    {\includegraphics[width=\panelW]{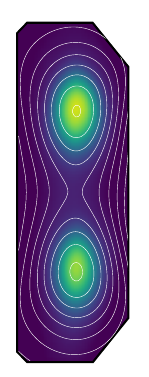}};
  \node[panel] (Ddpstd) at ($(Ddpmean.south east)+(\panelSep,0)$)
    {\includegraphics[width=\panelW]{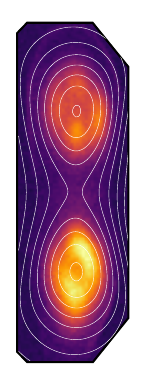}};
  \node[lab] at ($(Dgt.north)+(0,\labSep)$)          {$\vx_{_\text{GT}}$};
  \node[lab] at ($(Drdmap.north)+(0,\labSep-0.05)$)  {$\vx_{_\text{RD(MAP)}}$};
  \node[lab] at ($(Drdmean.north)+(0,\labSep)$)      {$\vmu_{_\text{RD}}$};
  \node[lab] at ($(Drdstd.north)+(0,\labSep)$)       {$\vsigma_{_\text{RD}}$};
  \node[lab] at ($(Ddpmean.north)+(0,\labSep)$)      {$\vmu_{_\text{DiffPIR}}$};
  \node[lab] at ($(Ddpstd.north)+(0,\labSep)$)       {$\vsigma_{_\text{DiffPIR}}$};
  \node[cbar] (Dcb1) at ($(Dgt.south)+(0,\cbarSep)$)
    {\includegraphics[width=\cbarW]{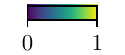}};
  \node[cbar] (Dcb2) at ($(Drdmap.south)+(0,\cbarSep)$)
    {\includegraphics[width=\cbarW]{\D/cbar_horiz_doublet.pdf}};
  \node[cbar] (Dcb3) at ($(Drdmean.south)+(0,\cbarSep)$)
    {\includegraphics[width=\cbarW]{\D/cbar_horiz_doublet.pdf}};
  \node[cbar] (Dcb4) at ($(Drdstd.south)+(0,\cbarSep)$)
    {\includegraphics[width=\cbarW]{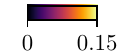}};
  \node[cbar] (Dcb5) at ($(Ddpmean.south)+(0,\cbarSep)$)
    {\includegraphics[width=\cbarW]{\D/cbar_horiz_doublet.pdf}};
  \node[cbar] (Dcb6) at ($(Ddpstd.south)+(0,\cbarSep)$)
    {\includegraphics[width=\cbarW]{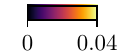}};

  \coordinate (row2) at ($(Dgt.south west |- Dcb1.south)-(0,\rowSep)$);

  \node[panel] (Sgt) at (row2)
    {\includegraphics[width=\panelW]{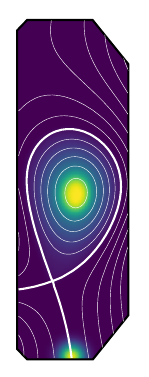}};
  \node[panel] (Srdmap) at ($(Sgt.south east)+(\groupSep,0)$)
    {\includegraphics[width=\panelW]{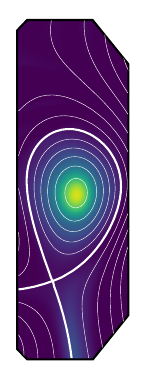}};
  \node[panel] (Srdmean) at ($(Srdmap.south east)+(\panelSep,0)$)
    {\includegraphics[width=\panelW]{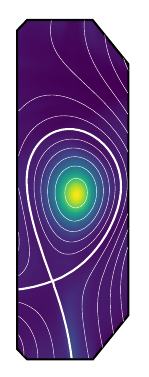}};
  \node[panel] (Srdstd) at ($(Srdmean.south east)+(\panelSep,0)$)
    {\includegraphics[width=\panelW]{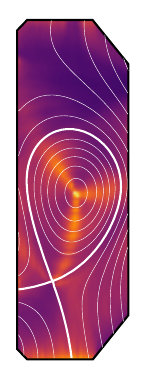}};
  \node[panel] (Sdpmean) at ($(Srdstd.south east)+(\groupSep,0)$)
    {\includegraphics[width=\panelW]{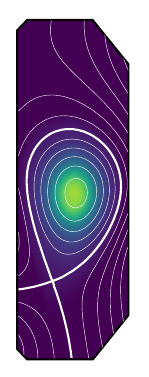}};
  \node[panel] (Sdpstd) at ($(Sdpmean.south east)+(\panelSep,0)$)
    {\includegraphics[width=\panelW]{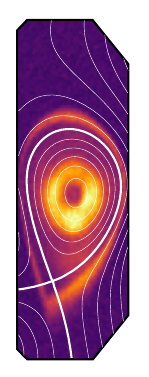}};
  \node[cbar] (Scb1) at ($(Sgt.south)+(0,\cbarSep)$)
    {\includegraphics[width=\cbarW]{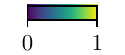}};
  \node[cbar] (Scb2) at ($(Srdmap.south)+(0,\cbarSep)$)
    {\includegraphics[width=\cbarW]{\D/cbar_horiz.pdf}};
  \node[cbar] (Scb3) at ($(Srdmean.south)+(0,\cbarSep)$)
    {\includegraphics[width=\cbarW]{\D/cbar_horiz.pdf}};
  \node[cbar] (Scb4) at ($(Srdstd.south)+(0,\cbarSep)$)
    {\includegraphics[width=\cbarW]{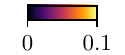}};
  \node[cbar] (Scb5) at ($(Sdpmean.south)+(0,\cbarSep)$)
    {\includegraphics[width=\cbarW]{\D/cbar_horiz.pdf}};
  \node[cbar] (Scb6) at ($(Sdpstd.south)+(0,\cbarSep)$)
    {\includegraphics[width=\cbarW]{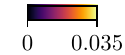}};

  \node[rowlab] at ($(Dgt.west)+(-\rowLabSep,0)$) {doublet};
  \node[rowlab] at ($(Sgt.west)+(-\rowLabSep,0)$) {strike point radiation};
\end{tikzpicture}
\captionsetup{justification=justified,singlelinecheck=false}
\caption{Out-of-distribution data: SXR reconstruction of synthetic profiles of a doublet plasma (top) and of a profile featuring radiation at the outer strike point (bottom). Left column: ground truth. Center: RD MAP estimate, posterior mean and standard deviation. Right: DiffPIR posterior mean and standard deviation.}
\label{fig:limitations}
\end{figure}

\hspace{-0.35cm}In the bottom row of Fig.\ref{fig:limitations}, we consider a different example, computing the SXR reconstructions of a synthetic emissivity profile that features, in addition to a Gaussian-like profile inside the core, a radiation blob localized at the outer divertor strike point.
Similar soft X-ray strike point radiation can be observed in TCV experiments.
In this case, DiffPIR disregards entirely the presence of the blob: the likelihood guidance is not sufficient to overcome the out-of-distribution character of the observed tomographic data.
The RD MAP estimate can be shown to behave similarly, if reconstruction is restricted to the core as done in all the phantom and experimental studies reported so far.
However, by removing this constraint, the RD estimates $\mathbf{x}_{_\text{RD(MAP)}}$ and $\vmu_{_\text{RD}}$ shown in Fig.\ref{fig:limitations} (bottom) reveal the presence of radiation outside the core.
The reconstruction quality remains unsatisfactory, as the anisotropic prior smooths the radiation along the divertor leg failing to correctly place it at the strike point.
This is due to the layout of the SXR diagnostic shown in Fig.\ref{fig:los_configurations}a, which features no horizontal lines-of-sight in the divertor region: the position of the radiation along the outer divertor leg is complex to determine.
Importantly, however, the presence of the radiation itself is correctly flagged, unlike in the DiffPIR case.
\\
Closer inspection of the DiffPIR posterior standard deviation, for both examples in Fig.\ref{fig:limitations}, reveals that $\vsigma_{_\text{DiffPIR}}$ appears noisier than in the cases analyzed in the previous sections.
This can be most clearly observed for the strike point radiation phantom, by zooming in on the darker region outside the core.
A less severe version of the same phenomenon can be observed also for the doublet, in the region comprised between the two lobes.
Moreover, in the strike point case, $\vsigma_{_\text{DiffPIR}}$ is non-negligible in the region outside the core; for the in-distribution data considered in Sections \ref{sxr_vs_dmpx}-\ref{experimental_data}, instead, $\vsigma_{_\text{DiffPIR}}$ quickly decays to values close to zero outside the core.
The posterior also fails to flag as significantly uncertain the strike point region.
These $\vsigma_{_\text{DiffPIR}}$ features are, in fact, a symptom of the fact that the model is confronted with out-of-distribution data.
In particular, the noisiness of $\vsigma_{_\text{DiffPIR}}$ is found to be a consequence of the noisiness of the posterior samples themselves: their inspection thus provides a valuable tool to diagnose the reliability of the DiffPIR reconstruction, as it helps flag potentially problematic cases.
\\
It is worth commenting further on the different behavior observed in the two examples.
While DiffPIR correctly recovers the doublet, it is blind to the strike point feature.
We ascribe this to the fact that the diffusion prior is encoded by a convolutional denoiser, which learns local structures (at multiple scales) rather than global plasma configurations.
While the doublet is, essentially, a rearrangement and combination of local structures that belong to the training samples, the strike point blob is a sharp and wall-adjacent local structure that was never seen by the network during its training.
Of course, strike point features could (and should) be added to the training set, if expected to be potentially present.
However, the described sensitivity of the diffusion model to out-of-distribution data remains.
\\
Ultimately, for a severely underdetermined reconstruction problem such as plasma tomography, the output of highly expressive models should always be compared with simpler algorithms such as RD, which are much less powerful but more interpretable and robust.
Tests designed to assess how a given algorithm would (fail to) reconstruct a specific emissivity feature, as in the discussed strike point example, can provide relevant information about the expected failure modes of a given algorithm.
These benchmarks and tests play a crucial role in the objective treatment of the uncertain nature of plasma tomography.

\section{Summary and Discussion}
\label{summary_and_discussion}
In this paper, we proposed a Bayesian data-driven approach to plasma tomography, based on score-based generative diffusion models.
The approach combines a physics-based likelihood with a prior learned from realistic emissivity profiles.
The prior is learned by training a U-Net convolutional neural network on a pretext denoising task, for a sequence of increasing noise levels.
Intuitively, a clean profile can be transformed into pure noise by injecting increasing amounts of noise according to the prescribed sequence.
The learned diffusion model, leveraging Tweedie's connection between denoising and the prior score $\nabla \log p(\mathbf{x})$, samples from the prior by iteratively denoising an initial pure noise realization: this corresponds to progressively reversing, step-by-step, the noise injection dynamics.
The learned prior can then be plugged into a given reconstruction problem, allowing sampling from the resulting posterior distribution.
\\
Here, we described the theory behind diffusion models and how they can be used for sampling.
In particular, we described the DDPM prior sampling scheme, and the DiffPIR algorithm for posterior sampling.
Moreover, we proposed DiffRD, a DiffPIR variant that combines the learned prior with the anisotropic equilibrium-informed RD prior introduced in Section \ref{inverse_problems_tomography}.
We learned a single diffusion prior for soft X-ray imaging at the TCV tokamak, from synthetic emissivity profiles.
Then, we conducted a phantom study to compare the diffusion and RD frameworks on three TCV soft X-ray diagnostics: SXR, DMPX, and Pilatus.
For the diffusion part, the trained network is deployed without adaptations for all three diagnostics, only adjusting the sampler hyperparameters once per diagnostic.
For DiffPIR, no reconstruction-specific hyperparameter tuning is needed, which is a considerable advantage compared to RD and DiffRD, whose anisotropic parameter $\alpha$ needs to be estimated for optimal performance.
\\
For SXR and DMPX, which are composed of multiple cameras with intersecting lines-of-sight, the equilibrium-free DiffPIR is found to outperform the equilibrium-informed RD in terms of reconstruction quality.
For the single-camera Pilatus, DiffPIR performs much worse than RD, as it fails to correctly localize the plasma position; the performance gap mostly fades if DiffPIR's reconstruction support is restricted to the core, while the equilibrium-informed DiffRD outperforms all other methods.
In all the considered diagnostics, for data consistent with the training distribution, the uncertainty estimated by the diffusion-based methods is found to provide meaningful information about the localization of the reconstruction error, and thus about the reliability of the model.
This is an improvement compared to RD, whose uncertainty carries little information about error localization.
In terms of empirical coverage, DiffPIR and RD provide useful (although typically conservative) operative credible intervals, while DiffRD tends to be overconfident about its own predictions.
We then applied DiffPIR to experimental SXR data; the obtained reconstructions are consistent with the magnetic equilibrium, despite not relying on it.
Moreover, we showed that inspection of the DiffPIR's posterior samples, which represent different plausible solutions to the observed data, can be used to assess how convincingly a reconstructed feature is supported by the data.
Finally, we discussed the limitations of diffusion models, whose performance degrades when confronted with data that significantly differ from the training samples.
Interestingly, however, DiffPIR's samples are found to be noisier for out-of-distribution data; while still failing to correctly localize the error, they tend to flag the reduced reliability of the model itself.
\\
The learned diffusion prior is more expressive than traditional hand-designed priors, and provides more informative uncertainty quantification.
The learning and reconstruction phases are decoupled: detectors found to be ill-behaved for a given discharge can simply be discarded, and the same model can be used, without retraining, for reconstruction from different diagnostics or to investigate upgraded diagnostic layouts.
This constitutes a clear advantage over end-to-end neural network approaches that learn a direct map from the data to the reconstructed profile, thus tying the learned model to a specific diagnostic layout; in addition, compared to PINN-based approaches, diffusion models remove the computational burden of training a new model for each considered time slice.
Moreover, the diffusion framework naturally provides uncertainty quantification, which was not available in previously proposed end-to-end deep learning approaches, and which, for PINNs, would require computationally expensive Bayesian formulations \cite{yang_bpinn_2021}.
Posterior inference based on diffusion models, while not real-time compatible, is computationally efficient: drawing $100$ samples takes $\approx 10$-$15\,\mathrm{s}$ on a single NVIDIA V100 GPU.
For diagnostics with sufficient line-of-sight coverage, such as SXR and DMPX, excellent performance is achieved even without relying on magnetic equilibrium information: this prevents propagating errors in the magnetic reconstruction into the emissivity estimation, and helps reduce the bias toward perfectly flux-aligned reconstructions.
Magnetic information remains essential for single-camera diagnostics such as Pilatus, where the hybrid DiffRD approach was found to well reconstruct 2D profiles, even in the presence of poloidal asymmetries; this comes, however, at the price of reintroducing the dependence on the magnetic reconstruction, and thus the drawbacks discussed above.
\\
The main limitation of the proposed approach is that, while more expressive, the learned diffusion priors are less robust and predictable than simpler hand-designed priors: model performance is sensitive to out-of-distribution data.
While the learned convolutional denoiser generalizes rather well to different combinations of known local structures, novel local structures might be discarded entirely.
Great care should thus be put into the construction of a representative training dataset, composed of diverse and physically realistic profiles.
Another source of concern is the scale of the data, as the learned prior is sensitive to it.
For data whose dynamic range is outside that of the training samples, normalization to the training scale was thus found to be necessary.
Under this normalization, DiffPIR degrades more than RD, although its performance margins remain significant.
Future research could investigate ways of improving robustness of the learned prior to scale variations.
Ultimately, it remains essential to benchmark the obtained reconstructions against those of simpler approaches, and to design specific tests to assess the expected failure modes of all deployed methods.
\\
A further limitation is that the learned prior is purely spatial, and reconstruction is performed frame by frame: unlike the time-resolved version of the PINN-based approach \cite{Rossi_2025}, the proposed diffusion framework does not exploit the temporal correlation of the emissivity.
Integrating regularity along an additional dimension, spatial or temporal, into diffusion-based reconstruction is an active research area \cite{chung3d, rozet2023}, and could be investigated for plasma tomography in future work.
\\
In this paper, we focused on soft X-ray emissivity reconstruction because a large number of physically realistic phantoms were immediately available, by the approach proposed in \cite{hamm2025}.
Future studies could apply the proposed diffusion framework to bolometry.
We view this as a very promising direction for future research: the enhanced expressiveness is expected to significantly improve the reconstruction of the emissivity features expected in the divertor, whose sharpness is often poorly reconstructed by hand-designed priors.
Application to bolometry, crucially, requires the creation of a large dataset of realistic emissivity profiles, which could rely on combining models of edge and core radiation, and potentially other expected features such as highly localized strike point radiation.
We encourage efforts toward open access sharing of the outputs of simulation codes: these practices would significantly accelerate the development of data-driven solutions for radiation studies.

\newpage

\section*{Acknowledgements}
This work was supported in part by the Swiss National Science Foundation, by the European Research Council (ERC-2020-AdG FunLearn-101020573), and by the Swiss National Science Foundation Sinergia Grant CRSII5 198569. This work has been carried out within the framework of the EUROfusion Consortium, via the Euratom Research and Training Programme (Grant Agreement No 101052200 — EUROfusion) and funded by the Swiss State Secretariat for Education, Research and Innovation (SERI). Views and opinions expressed are however those of the author(s) only and do not necessarily reflect those of the European Union, the European Commission, or SERI. Neither the European Union nor the European Commission nor SERI can be held responsible for them.

\newpage

\bibliographystyle{unsrt}
\typeout{}
\bibliography{diffusion_tomo_biblio}

\newpage

\appendix

\section{Denoising Diffusion Probabilistic Models: the DDPM sampling algorithm}\label{app:sampling}
The DDPM forward process is a Markov chain of Gaussian transitions, so the conditional distribution of $\mathbf{x}_{t-1}$ given both $\mathbf{x}_t$ and the clean profile $\mathbf{x}_0$ is itself Gaussian \cite{ho_2020},
\begin{equation}\label{ddpm_posterior}
    q(\mathbf{x}_{t-1}\,\vert\,\mathbf{x}_t, \mathbf{x}_0)
    = \mathcal{N}\!\Big(
      \tfrac{\sqrt{\bar{\alpha}_{t-1}}\,\beta_t}{1-\bar{\alpha}_t}\,\mathbf{x}_0
      + \tfrac{\sqrt{\alpha_t}\,(1-\bar{\alpha}_{t-1})}{1-\bar{\alpha}_t}\,\mathbf{x}_t,\;
      \tilde{\beta}_t\,\mathbf{I}\Big),
    \qquad
    \tilde{\beta}_t = \frac{1-\bar{\alpha}_{t-1}}{1-\bar{\alpha}_t}\,\beta_t\;,
\end{equation}
with $\alpha_t = 1-\beta_t$ and $\bar{\alpha}_0 = 1$.
Its mean interpolates between the current iterate $\mathbf{x}_t$ and the clean profile.
Since $\mathbf{x}_0$ is unknown at sampling time, DDPM replaces it with the denoiser's MMSE estimate $\hat{\mathbf{x}}_0(\mathbf{x}_t)$, obtained from the noise prediction $\boldsymbol{\epsilon}_\theta$ by inverting the corruption $\mathbf{x}_t = \sqrt{\bar{\alpha}_t}\,\mathbf{x}_0 + \sqrt{1-\bar{\alpha}_t}\,\mathbf{z}$.
This yields the ancestral sampling scheme of Algorithm~\ref{alg:ddpm_sampling}, which is a discretization of the reverse-time SDE of Eq.~\eqref{reverse_sde} \cite{song_2021}.
By substituting the expression of $\mathbf{x}_t$ computed from line 4 of the algorithm (which yields $\mathbf{x}_t=\sqrt{\bar{\alpha}_t}\hat{\mathbf{x}}_0+\sqrt{1-\bar{\alpha}_t}\hat{\mathbf{z}}$) into line 5, it can be seen that the DDPM update in line 7 and the DiffPIR update in line 8 of Algorithm \ref{alg:diffrd}, while written in different variables, are actually of the same type.
\begin{algorithm}
\caption{DDPM ancestral sampling from the learned prior}
\label{alg:ddpm_sampling}
\begin{algorithmic}[1]
\State $\mathbf{x}_T \sim \mathcal{N}(\mathbf{0},\mathbf{I})$
\For{$t = T, \ldots, 1$}
    \State $\hat{\mathbf{z}} \gets \boldsymbol{\epsilon}_\theta(\mathbf{x}_t, t)$
        \Comment{predicted noise}
    \State $\hat{\mathbf{x}}_0 \gets \big(\mathbf{x}_t - \sqrt{1-\bar{\alpha}_t}\,\hat{\mathbf{z}}\big)/\sqrt{\bar{\alpha}_t}$
        \Comment{MMSE estimate of the clean profile}
    \State $\boldsymbol{\mu}_t \gets
        \frac{\sqrt{\bar{\alpha}_{t-1}}\,\beta_t}{1-\bar{\alpha}_t}\,\hat{\mathbf{x}}_0
        + \frac{\sqrt{\alpha_t}\,(1-\bar{\alpha}_{t-1})}{1-\bar{\alpha}_t}\,\mathbf{x}_t$
        \Comment{mean of Eq.~\eqref{ddpm_posterior} with $\mathbf{x}_0 \to \hat{\mathbf{x}}_0$}
    \State $\mathbf{z} \sim \mathcal{N}(\mathbf{0},\mathbf{I})$ \textbf{if} $t>1$ \textbf{else} $\mathbf{z} \gets \mathbf{0}$
    \State $\mathbf{x}_{t-1} \gets \boldsymbol{\mu}_t + \sqrt{\tilde{\beta}_t}\,\mathbf{z}$
\EndFor
\State \Return $\mathbf{x}_0$
\end{algorithmic}
\end{algorithm}

\end{document}